\documentclass[aps,prb,twocolumn]{revtex4}
\usepackage{graphicx}
\usepackage{dcolumn}
\usepackage{amsmath}
\usepackage{amssymb}
\usepackage{float}
\usepackage{epstopdf}
\usepackage{nicefrac}

\def\rv{{\bf r}}
\def\fv{{\bf f}}
\def\sv{{\bf s}}

\def\beq{\begin{equation}}
\def\eeq{\end{equation}}

\def\enerdens{w}

\begin{document}
\title{Energy densities in the strong-interaction limit of density functional theory}
\author{Andr\'e Mirtschink,$^1$ Michael Seidl,$^2$ and Paola Gori-Giorgi$^1$}
\affiliation{$^1$Department of Theoretical Chemistry and Amsterdam Center for Multiscale Modeling, FEW, Vrije Universiteit, De Boelelaan 1083, 1081HV Amsterdam, The Netherlands\\
$^2$Institute of Theoretical Physics, University of Regensburg, D-93040 Regensburg, Germany}
\date{\today}

\begin{abstract}
We discuss energy densities in the strong-interaction limit of density functional theory, deriving an exact expression  within the definition (gauge) of the electrostatic potential of the exchange-correlation hole. Exact results for small atoms and small model quantum dots are compared with available approximations defined in the same gauge. The idea of a local interpolation along the adiabatic connection is discussed, comparing the energy densities of the Kohn-Sham, the physical, and the strong-interacting systems. We also use our results to analyze the local version of the Lieb-Oxford bound, widely used in the construction of approximate exchange-correlation functionals. 
\end{abstract}

\maketitle
\section{Introduction}

Increasing the accuracy of the approximations to the exchange-correlation energy functional $E_{xc}[\rho]$ of Kohn-Sham (KS) density functional theory (DFT) is of crucial importance for research areas ranging from theoretical chemistry and biochemistry to solid-state and surface physics (for a recent review, see, e.g., Ref.~\onlinecite{CohMorYan-CR-12}). 

A piece of {\em exact} information on $E_{xc}[\rho]$ is provided by the strong-interaction limit of DFT, in which the coupling constant of the electron-electron interaction becomes infinitely large while the one-electron density $\rho(\rv)$ does not change.\cite{Sei-PRA-99,SeiPerLev-PRA-99,SeiPerKur-PRA-00,SeiPerKur-PRL-00} This defines a fictitious system with the same density of the physical one and maximum possible correlation between the relative electronic positions, useful to describe situations in which restricted Kohn-Sham DFT encounters problems, such as low-density many-particle scenarios and the breaking of the chemical bond.\cite{GorSeiVig-PRL-09,GorSei-PCCP-10} The exact mathematical structure of this limit has been uncovered only recently,\cite{SeiGorSav-PRA-07,GorVigSei-JCTC-09,ButDepGor-PRA-12} and exact calculations (at least for simple systems) have started to become available.\cite{GorSeiVig-PRL-09,GorSei-PCCP-10,RasSeiGor-PRB-11} 

The aim of this paper is to make a step forward in the inclusion of this new piece of exact information into approximations to the exchange-correlation energy functional $E_{xc}[\rho]$. Previous attempts in this direction\cite{SeiPerLev-PRA-99,SeiPerKur-PRA-00,SeiPerKur-PRL-00} focused on global (i.e., integrated over all space) quantities,  introducing size-consistency errors. The exact solution of the strong-interaction limit, now available, makes accessible not only global, but also {\em local} quantities, from which it is easier to construct size-consistent approximations\cite{Bec-JCP-05,BecJoh-JCP-07,PerStaTaoScu-PRA-08} (for a critical review on size consistency of approximate energy density functionals see, e.g.,  Ref.~\onlinecite{GorSav-JPCS-08} and, especially, Ref.~\onlinecite{Sav-CP-09}). 

Local quantities are in general not uniquely defined (for a complete discussion see, e.g., Ref.~\onlinecite{BurCruLam-JCP-98}). Here we focus on the conventional, physical transparent, definition in terms of the electrostatic energy of the exchange-correlation hole. We derive the exact expression for this energy density in the strong-interaction limit, and we evaluate it for small atoms and small model quantum dots, making comparisons with available approximations within the same definition (same {\em gauge}).
We then discuss the idea of a local interpolation along the adiabatic connection by comparing energy densities in the physical case, the weak- and the strong-interaction limit. 

As a byproduct, our results allow to analyze the local version of the Lieb-Oxford bound, a condition widely used to construct approximate exchange-correlation functionals. As well known, the Lieb-Oxford bound is an exact condition\cite{Lie-PLA-79,LieOxf-IJQC-81,ChaHan-PRA-99} on the global $E_{xc}[\rho]$. Many non-empirical approximate functionals, however, use its local version, which is a sufficient but not necessary condition to ensure the global bound (see, e.g., Refs.~\onlinecite{PerBurErn-PRL-96,TaoPerStaScu-PRL-03} and~\onlinecite{HauOdaScuPerCap-JCP-12}). Our analysis strongly suggests that the local version of the Lieb-Oxford bound is formulated in the gauge of the electrostatic potential of the exchange-correlation hole. This, in turn, implies that the local bound is certainly violated at least in the tail region of a molecular or atomic density, and in the bond region of a stretched molecule.

The paper is organized as follows. In the next Sec.~\ref{sec_inter} we review the DFT adiabatic connection as a tool to build approximate $E_{xc}[\rho]$, highlighting the role of the strong-interaction limit and discussing the size-consistency problem of interpolations based on global quantities. In Sec.~\ref{sec_energydens} we discuss energy densities in general and we introduce the gauge of the electrostatic energy associated to the exchange-correlation hole. The exact expression for this quantity in the strong-interaction limit is derived in Sec.~\ref{sec_strong}, where also approximations are discussed. In particular, we analyze the ``point-charge plus continuum'' (PC) model functional,\cite{SeiPerKur-PRA-00} showing that it is meant to be an approximation to the energy density within the same conventional definition considered here. Energy densities along the adiabatic connection are discussed and analyzed in Sec.~\ref{sec_enedenslambda}. We then use our results to discuss the local version of the Lieb-Oxford bound in Sec.~\ref{sec_LO}. The last Sec.~\ref{sec_conc} is devoted to conclusions and perspectives. Finally, a simple illustration for the case of the uniform electron gas is reported in the Appendix.

\section{Interpolation along the adiabatic connection}

\label{sec_inter}

Within the framework of the adiabatic connection\cite{HarJon-JPF-74, LanPer-SSC-75, GunLun-PRB-76} the exchange-correlation energy can be expressed by the coupling constant integration
	\begin{align}
		E_{xc}[\rho]&=\int_0^1 d\lambda\big\langle\Psi_\lambda[\rho]\big\vert\hat{V}_{ee}\big\vert\Psi_\lambda[\rho]\big\rangle-U[\rho]\label{eq:cci}\\
			&=:\int_0^1 d\lambda\; W_\lambda[\rho],
	\end{align}
with $\Psi_\lambda[\rho]$ being the ground state wave function of a fictitious system with scaled electron-electron interaction
	\begin{align}
		\hat{H}=\hat{T} +\lambda\hat{V}_{ee} +\hat{V}_{ext}^\lambda.
		\label{eq:Hlambda}
	\end{align}
The external potential $\hat{V}_{ext}^\lambda$ is adjusted to keep the density $\rho_\lambda(\rv)$ out of $\Psi_\lambda[\rho]$ in agreement with the physical density, $\rho_\lambda(\rv)=\rho_{1}(\rv)\equiv \rho(\rv)$. For the weak interaction limit ($\lambda=0$) we encounter the Kohn-Sham\cite{KohSha-PR-65} reference system and the integrand $W_0[\rho]$ becomes the exact exchange energy $E_x[\rho]$. For the strong-interaction limit ($\lambda\rightarrow\infty$) a reference system within the strictly correlated electrons concept (SCE)\cite{SeiPerLev-PRA-99, Sei-PRA-99, SeiGorSav-PRA-07,GorVigSei-JCTC-09} can be defined. The asymptotic expansions of $W_{\lambda}[\rho]$ are
	\begin{align}
		W_{\lambda\rightarrow 0}[\rho]=E_x[\rho]+2\,\lambda\, E_c^{\rm GL2}[\rho]+O(\lambda^2)\label{eq:expl0}\\
		W_{\lambda\rightarrow\infty}[\rho]=W_\infty[\rho]+\frac{W'_\infty[\rho]}{\sqrt{\lambda}}+O(\lambda^{-p}),\label{eq:explinf}
	\end{align}
where $E_c^{\rm GL2}[\rho]$ is the correlation energy given by second order G\"{o}rling-Levy perturbation theory (GL2)\cite{GorLev-PRB-93} and  $p\geq 5/4$.\cite{GorVigSei-JCTC-09} Exact expressions for the functionals $W_\infty[\rho]$ and $W'_\infty[\rho]$ are given, respectively, in Refs.~\onlinecite{SeiGorSav-PRA-07} and \onlinecite{GorVigSei-JCTC-09}. 

Expression \eqref{eq:cci} for the exchange-correlation energy is exact as long as the exact dependence of the integrand on $\lambda$ is known.\cite{TeaCorHel-JCP-09,TeaCorHel-JCP-10} As this is obviously not the case, Eq. \eqref{eq:cci} still enables approximate exchange-correlation energies by modeling $W_\lambda[\rho]$ along the adiabatic connection.

Attempts towards approximate $W_\lambda[\rho]$ have been undertaken by Becke,\cite{Bec-JCP-93a} introducing the half and half functional. A model is defined assuming a linear dependence of $W_\lambda[\rho]$ on $\lambda$ and setting $W_0[\rho]$ equal to exact exchange and $W_1[\rho]$ to LSDA exchange-correlation. This results in a functional with 50\% exact exchange and 50\% LSDA exchange-correlation. Further adjustment of the portion of exact exchange by semi-empirical arguments gives rise to hybrid functionals like B3LYP.\cite{Bec-JCP-93, SteDevChaFri-JPC-94, LeeYanPar-PRB-88, VosWilNus-CJP-80} The adiabatic connection may also be used for the construction of non-empirical hybrids as in Ref.~\onlinecite{BurErnPer-CPL-97}. Here a model for $W_\lambda[\rho]$ is defined consisting of two intersected straight lines fixed by exact exchange, GGA exchange and GGA exchange-correlation. 
Ernzerhof\cite{Ern-CPL-96} introduced a curved model by proposing a Pad\'{e} interpolation for the integrand using as input exact exchange and $E_c^{\rm GL2}[\rho]$ in the weak interaction limit and GGA exchange-correlation for $\lambda=1$.

The mentioned models for the integrand (i.e. all except B3LYP) have in common that for the weak interaction limit exact exchange is employed and for the physical situation with $\lambda=1$ information from approximate DFT (DFA) is used. The argument for the recourse to exact exchange is that DFA exchange works well only if combined with DFA correlation. This is due to error cancellation. Consequently, DFA exchange-correlation can be used for the physical case where exchange and correlation are employed together. As error cancellation in DFA exchange-correlation might not be satisfactory, a continuation of the ansatz of Ernzerhof\cite{Ern-CPL-96} is possible by taking DFA exchange-correlation at some intermediate $\lambda$ instead of $\lambda=1$. This would allow to balance the exchange error with the correlation error. Along this line, Mori-S\'{a}nchez, Cohen and Yang\cite{MorCohYan-JCPa-06} constructed their MCY1 functional: a Pad\'{e} interpolation is undertaken with exact exchange and meta-GGA exchange input in the weak interaction limit and meta-GGA exchange-correlation for an intermediate $\lambda$ (chosen semi-empirically). 

The discussed models clearly outperform the stand alone DFAs they are based on.\cite{Bec-JCP-93a,BurErnPer-CPL-97,Ern-CPL-96,MorCohYan-JCPa-06} Nonetheless, employment of DFA quantities in their construction can lead to serious misbehavior in the curvature of the integrand as demonstrated by Peach, Teale and Tozer\cite{PeaTeaToz-JCP-07} by comparison of the MCY1 approximation with accurate quantities along the adiabatic connection (see, e.g., Fig. 3 in Ref.~\onlinecite{PeaTeaToz-JCP-07}). In the same paper the authors show that accurate exchange-correlation can be recovered by establishing an interpolation with accurate full-CI ingredients.

A model that avoids unfavorable DFA bias is the interaction strength interpolation (ISI).\cite{SeiPerLev-PRA-99, Sei-PRA-99, SeiPerKur-PRA-00, GorVigSei-JCTC-09} Here information from the weak interaction limit is employed, namely exact exchange and GL2, together with information from the strong interacting limit through the point-charge plus continuum (PC) model, which provides approximate expressions for $W_\infty[\rho]$ and $W'_\infty[\rho]$. The $\lambda$ dependence of $W_\lambda[\rho]$ is then modeled by an interpolation between the two limits. Nowadays, the functionals $W_\infty[\rho]$ and $W'_\infty[\rho]$ can be  accurately computed within the SCE many-electron formalism.\cite{SeiGorSav-PRA-07, GorVigSei-JCTC-09} Refs.~\onlinecite{SeiGorSav-PRA-07} and \onlinecite{GorVigSei-JCTC-09} compare the PC solutions with the exact SCE values for small atoms: while  $W^{\rm PC}_\infty[\rho]$ is a very reasonable approximation to its exact counterpart,\cite{SeiGorSav-PRA-07} the original $W'^{\rm PC}_\infty[\rho]$ turned out to be much less accurate.\cite{GorVigSei-JCTC-09} The exact results could be used to propose a revised PC approximation $W'^{\rm revPC}_\infty[\rho]$ having accuracy similar to the one of $W^{\rm PC}_\infty[\rho]$. Further comparison is undertaken in section \ref{sec:pc} of this paper for a more refined quantity, the local energy density as it will be defined in the next sections.  

Although unplugged from any DFA bias (if we use exact input quantities), an unpleasant feature of the ISI is the violation of size consistency. This is due to the non-linear way the (size-consistent) ingredients $W_0[\rho]$, $W'_0[\rho]$, $W_\infty[\rho]$ and $W'_\infty[\rho]$ enter the interpolation. For example, the revised ISI (which behaves better in the $\lambda\to\infty$ limit than the original ISI) reads\cite{GorVigSei-JCTC-09} 
	\begin{align}
		W_\lambda^{\rm revISI}[\rho]=\frac{\partial}{\partial\lambda}\left(a[\rho]\lambda+\frac{b[\rho]\lambda}{\sqrt{1+c[\rho]\lambda}+d[\rho]}\right)
	\end{align}
where $a$, $b$, $c$ and $d$ are non linear functions of $W_0[\rho]$, $W'_0[\rho]$, $W_\infty[\rho]$ and $W'_\infty[\rho]$, determined by imposing the asymptotic expansions of Eqs.~\eqref{eq:expl0} and \eqref{eq:explinf}:
\begin{eqnarray}
	a[\rho] & = & W_{\infty}[\rho]\\
	b[\rho] & = & -\frac{8\,E_c^{\rm GL2}[\rho] W'_{\infty}[\rho]^2}{(E_x[\rho]-W_{\infty}[\rho])^2} \\
	c[\rho] & = & \frac{16\,E_c^{\rm GL2}[\rho]^2 W'_{\infty}[\rho]^2}{(E_x[\rho]-W_{\infty}[\rho])^4} \\
	d[\rho] & = & -1-\frac{8\,E_c^{\rm GL2}[\rho] W'_{\infty}[\rho]^2}{(E_x[\rho]-W_{\infty}[\rho])^3}.
\end{eqnarray}
Notice that the lack of size consistency is shared by all functionals in which the exact exchange energy (or any global energy) enters in a nonlinear way. Thus also, for example, MCY1.

As a final remark on the revISI functional, we can add that if one makes the approximation $E_c^{\rm GL2}\approx E_c^{\rm MP2}$, it can be viewed as a double hybrid functional (see, e.g., Refs.~\onlinecite{Gri-JCP-06,ShaTouSav-JCP-11,BreAda-JCP-11,TouShaBreAda-JCP-11}). With respect to available double hybrids, the revISI lacks size consistency, but it has the advantage of being able to deal with the small-gap systems problematic for perturbation theory. The practical impact of the lack of size consistency of the revISI functional still needs to be tested.

\section{Energy densities: definitions}
\label{sec_energydens}
A possible way to recover size consistency in the ISI framework is to use a {\em local} integrand in Eq.~\eqref{eq:cci}:
	\begin{align}
		E_{xc}[\rho]=\int_0^1 d\lambda \int d \rv \,\rho(\rv)\,\enerdens_\lambda[\rho](\rv).
		\label{eq:enrgdens}
	\end{align}
The idea is then to build a local model, $\enerdens^{\rm ISI}_\lambda[\rho](\rv)$, by interpolating between the $\lambda\to 0$ and the $\lambda\to\infty$ limits. As said, the energy density $\enerdens_\lambda[\rho](\rv)$ is not uniquely defined, so that an important requirement here is that the input local quantities in the weak- and in the strong-interaction limits are defined in the same way (same gauge). 

One of the most widely used definitions of the energy density in DFT is in terms of the exchange-correlation hole (see, e.g., Refs.~\onlinecite{Bec-JCP-05,BecJoh-JCP-07} and \onlinecite{PerStaTaoScu-PRA-08}) $h_{xc}^{\lambda}(\rv,\rv')$, 
\begin{equation}
	\enerdens_\lambda[\rho](\rv)=\frac{1}{2}\int \frac{h_{xc}^\lambda(\rv,\rv')}{|\rv-\rv'|}\,d\rv',
	\label{eq:defenergydensity}
\end{equation}
where
\begin{equation}
h_{xc}^{\lambda}(\rv,\rv')=\frac{P_2^\lambda(\rv,\rv')}{\rho(\rv)}-\rho(\rv'),	
\label{eq_hxclambda}
\end{equation}
and the pair-density $P_2^\lambda(\rv,\rv')$ is obtained from the wave function $\Psi_{\lambda}[\rho]$ of Eqs.~\eqref{eq:cci}-\eqref{eq:Hlambda}:
\begin{eqnarray}
 & & P_2^\lambda(\rv,\rv')= N(N-1)\times \nonumber \\
  & & \sum_{\sigma_1...\sigma_N}\int|\Psi_\lambda(\rv\sigma_1,\rv'\sigma_2,\rv_3\sigma_3,...,\rv_N\sigma_N)|^2d\rv_3...d\rv_N.\qquad
\label{eq_P2}
\end{eqnarray}
In the definition of Eq.~\eqref{eq:defenergydensity}, $\enerdens_\lambda[\rho](\rv)$ is the electrostatic potential of the exchange-correlation hole (a negative charge distribution normalized to $-1$), around a reference electron in $\rv$. This quantity at the physical coupling strength $\lambda=1$ (plus the Hartree potential)  has been also called $v_{\rm cond}(\rv)$ in the literature (see, e.g., Ref.~\onlinecite{BuiBaeSni-PRA-89}).

The energy density $\enerdens_\lambda[\rho](\rv)$ in the $\lambda\to\infty$ limit in the gauge of Eq.~\eqref{eq:defenergydensity} is the central quantity of this paper: we will derive in the next section an exact expression using the strictly-correlated electron concept, and we will evaluate it for small atoms and quantum dots. Notice that the relevance of  $\enerdens_\infty[\rho](\rv)$ for constructing a new generation of approximate $E_{xc}[\rho]$ has also been pointed out very recently by Becke.\cite{Bec-cecam-11}

\section{Energy densities in the strong interaction limit}
\label{sec_strong}
\subsection{Exact}\label{sec:SCEexact}
When $\lambda\to\infty$ the wave function $\Psi_\lambda[\rho]$ tends to the strictly-correlated electron state, $\Psi_{\lambda\to\infty}[\rho]\to\Psi_{\rm SCE}[\rho]$, with\cite{SeiGorSav-PRA-07,GorVigSei-JCTC-09} 
\begin{eqnarray}
& & |\Psi_{\rm SCE}(\rv_1,...,\rv_N)|^2  =  \frac{1}{N!}\sum_{\mathcal{P}}\int d\sv \frac{\rho(\sv)}{N}
\delta(\rv_1-\fv_{\mathcal{P}(1)}(\sv)) \nonumber \\
& &  \times \delta(\rv_2-\fv_{\mathcal{P}(2)}(\sv))...\delta(\rv_N-\fv_{\mathcal{P}(N)}(\sv)),
\label{eq_PsiSCE}
\end{eqnarray}
where $\fv_1,..,\fv_N$ are ``co-motion functions'', with $\fv_1(\rv)\equiv\rv$, and $\mathcal{P}$ denotes a permutation of $\{1,...N\}$. This means that the $N$ points $\rv_1,...,\rv_N$ in 3D space found upon simultaneous measurement of the $N$ electronic positions in the SCE state  always obey the
$N-1$ relations
\beq
\rv_i=\fv_i(\rv_1)\qquad(i=2,...,N).
\label{cofs}
\eeq
In other words, the position of one electron determines all the relative $N-1$ electronic positions (limit of {\em strict correlation}).
All the $N-1$ co-motion functions $\fv_i(\sv)$  satisfy the differential equation
\beq
\rho(\fv_i(\rv))d \fv_i(\rv)=\rho(\rv)d \rv,
\label{eq_fdiff}
\eeq
which, together with the group properties\cite{SeiGorSav-PRA-07,ButDepGor-PRA-12}  of the $\fv_i(\rv)$, ensure that the SCE wavefunction of Eq.~(\ref{eq_PsiSCE}) yields the given density $\rho(\rv)$. Equation \eqref{eq_fdiff} has also a simple physical interpretation: since the position of one electron determines the position of all the others, the probability of finding one electron in the volume element $d\rv$ about $\rv$ must be the same of finding the $i^{\rm th}$ electron in the volume element $d\fv_i(\rv)$ about $\fv_i(\rv)$. 

Inserting Eq.~\eqref{eq_PsiSCE} into Eq.~\eqref{eq_P2} we obtain for the pair density $P_2^{\lambda\to\infty}(\rv_1,\rv_2)=P_2^{\rm SCE}(\rv_1,\rv_2)$ in the strong-interaction limit
\beq 
P_2^{\rm SCE}(\rv_1,\rv_2)=\sum_{\substack{i,j=1 \\ i\neq j}}^N\int d\sv \frac{\rho(\sv)}{N}\delta(\rv_1-\fv_i(\sv))\delta(\rv_2-\fv_j(\sv)),
\label{eq_P2SCE}
\eeq
which also has a transparent physical meaning: two electrons can only be found at strictly correlated relative positions.

We first compute
\beq
	\int \frac{P_2^{\rm SCE}(\rv,\rv')}{|\rv-\rv'|}d\rv'=\sum_{\substack{i,j=1 \\ i\neq j}}^N\int d\sv\frac{\rho(\sv)}{N}\frac{\delta(\rv-\fv_i(\sv))}{|\rv-\fv_j(\sv)|},
	\label{eq_eleSCE1}
\eeq
where, in the right-hand-side, we have already integrated over the variable $\rv'$. From the properties of the Dirac delta distribution  and of the co-motion functions, Eq.~\eqref{eq_eleSCE1} becomes
\begin{eqnarray}
	\int \frac{P_2^{\rm SCE}(\rv,\rv')}{|\rv-\rv'|}d\rv' & = & \frac{1}{N}\sum_{\substack{i,j=1 \\ i\neq j}}^N \frac{\rho\left(\fv_i^{-1}(\rv)\right)|{\rm det}\, \partial_\alpha f_{i,\beta}^{-1}(\rv)|}{|\rv-\fv_j\left(\fv_i^{-1}(\rv)\right)|} \nonumber \\
& = & \frac{\rho(\rv)}{N}\sum_{\substack{i,j=1 \\ i\neq j}}^N \frac{1}{|\rv-\fv_j\left(\fv_i^{-1}(\rv)\right)|},
\label{eq_doublesum}
\end{eqnarray}
where $|{\rm det}\, \partial_\alpha f_{i,\beta}^{-1}(\rv)|$ (with $\alpha,\beta=x,y,z$) is the determinant of the Jacobian of the transformation $\rv\to\fv_i^{-1}(\rv)$, and we have used the fact that all the $\fv_i(\rv)$ (and their inverses, which, by virtue of the group properties of the co-motion functions are also co-motion functions for the same configuration \cite{SeiGorSav-PRA-07,GorVigSei-JCTC-09}) satisfy Eq.~\eqref{eq_fdiff}. Now we can use once more the group properties of the co-motion functions to recognize that for all $i\neq j$ the function $\fv_j\left(\fv_i^{-1}(\rv)\right)$ must be another co-motion function with the exclusion of $\fv_1(\rv)= \rv$ (the identity can arise only if $i=j$). The double sum in the last term of Eq.~\eqref{eq_doublesum} is then exactly equal to $N$ times a single sum over all the co-motion functions $\fv_k(\rv)$ with $k\ge 2$, so that
\beq
\int \frac{P_2^{\rm SCE}(\rv,\rv')}{|\rv-\rv'|}d\rv'=\rho(\rv)\sum_{k=2}^N\frac{1}{|\rv-\fv_k(\rv)|}.
\label{eq_vcondinf}
\eeq
Inserting Eq.~\eqref{eq_vcondinf} into Eqs.~\eqref{eq:defenergydensity}-\eqref{eq_hxclambda} we finally obtain
\beq
\enerdens_\infty(\rv)=\frac{1}{2}\sum_{k=2}^N\frac{1}{|\rv-\fv_k(\rv)|}-\frac{1}{2}v_H(\rv),
\label{eq_endensSCE}
\eeq
where $v_H(\rv)$ is the Hartree potential. Notice that in previous work the exact $W_\infty[\rho]$ was given as\cite{SeiGorSav-PRA-07,GorVigSei-JCTC-09,GorSei-PCCP-10}
\beq
W_\infty[\rho]=\frac{1}{2}\int d\rv\, \frac{\rho(\rv)}{N} \, \sum_{\substack{i,j=1 \\ i\neq j}}^N \frac{1}{|\fv_i(\rv)-\fv_j(\rv)|}-U[\rho],
\eeq
suggesting a corresponding energy density
\beq
\tilde{\enerdens}_\infty(\rv)=\frac{1}{N}\sum_{i=1}^N\left(\frac{1}{2}\sum_{\substack{j=1 \\ j\neq i}}^N \frac{1}{|\fv_i(\rv)-\fv_j(\rv)|}-\frac{1}{2}v_H(\fv_i(\rv))\right).
\label{eq_endensSCEold}
\eeq
Equations \eqref{eq_endensSCEold} and \eqref{eq_endensSCE} yield the same $W_\infty[\rho]$ when integrated with the density $\rho(\rv)$, but are locally different. They show a general feature of the co-motion functions: any given energy density $\enerdens_\infty^a(\rv)$ can be always transformed into a different energy density $\enerdens_\infty^b(\rv)$ defined as
\beq
\enerdens_\infty^b(\rv)=\frac{1}{N}\sum_{i=1}^N \enerdens_\infty^a(\fv_i(\rv)).
\label{eq_changegauge}
\eeq
When multiplied by the density $\rho(\rv)$, $\enerdens_\infty^a(\rv)$ and $\enerdens_\infty^b(\rv)$ integrate to the same quantity, because all the co-motion functions (and their inverses, which are also co-motion functions) satisfy Eq.~\eqref{eq_fdiff}. Only Eq.~\eqref{eq_endensSCE} corresponds to the gauge of the exchange-correlation hole defined by Eqs.~\eqref{eq:defenergydensity}-\eqref{eq_P2}.

\subsection{Approximations: the PC model}\label{sec:pc}
The point-charge-plus-continuum (PC) model\cite{Ons-JPC-39,SeiPerKur-PRA-00} is a physically sound approximation to the $\lambda\to\infty$ indirect electron-electron repulsion energy $W_\infty[\rho]$. The idea is to rewrite the indirect Coulomb interaction energy $W_\lambda[\rho]$ as the electrostatic energy $E_{es}[\Psi_\lambda,\rho]$ of a system of $N$ electrons in the state $\Psi_\lambda[\rho]$ embedded in a smeared background of positive charge $\rho_+(\rv)=\rho(\rv)$.\cite{SeiPerKur-PRA-00} In fact, this total electrostatic energy $E_{es}[\Psi_\lambda,\rho]$ is just the sum of the electron-electron repulsion energy, $E_{ee}=\langle\Psi_\lambda|\hat{V}_{ee}|\Psi_\lambda\rangle$, the electron-background attraction energy, $E_{eb}=-2U[\rho]$, and the background-background repulsion energy $E_{bb}=U[\rho]$, thus yielding exactly $E_{es}[\Psi_\lambda,\rho]=E_{ee}+E_{eb}+E_{bb}= W_\lambda[\rho]$. 

This relation is valid for every $\lambda$, but in the $\lambda\to\infty$ limit, when $\Psi_\lambda\to\Psi_{\rm SCE}$, we expect that the electrons minimize $E_{es}[\Psi_\lambda,\rho]$ by occupying relative positions that divide the space into neutral cells with possibly zero (or weak) lowest-order electrostatic multipole moments.\cite{SeiPerKur-PRA-00} The idea is then that for one of the SCE configurations $\{\rv,\fv_2(\rv),...,\fv_N(\rv)\}$ we may approximate the indirect electron-electron repulsion  by the sum of the electrostatic energies of all the cells (i.e., we neglect the cell-cell interaction in view of their neutrality and low multipole moments):
\beq
\epsilon_{es}(\rv,\fv_2(\rv),...,\fv_N(\rv))\approx\sum_{i=1}^N E_{\rm cell}([\rho];\fv_i(\rv)),
\label{eq_esofr}
\eeq
where $E_{\rm cell}([\rho];\rv_i)$ is the electrostatic energy of the cell around an electron at position $\rv_i$, equal to the sum of the attraction between the electron and the background contained in the cell and the background-background repulsion inside the cell.\cite{SeiPerKur-PRA-00}

Notice that, for a given SCE configuration, the electrostatic energy $\epsilon_{es}(\rv,\fv_2(\rv),...,\fv_N(\rv))$  of Eq.~\eqref{eq_esofr}  is equal to $N\tilde{\enerdens}_\infty(\rv)$, where $\tilde{\enerdens}_\infty(\rv)$ is given in Eq.~\eqref{eq_endensSCEold}. The PC model is then trying to approximate $N\tilde{\enerdens}_\infty(\rv)$ by constructing the electrostatic energy $E_{\rm cell}([\rho];\rv_i)$ of a cell around the electron at position $\rv_i$. However, and this is a crucial step to understand the gauge of the PC model, once an approximation for $E_{\rm cell}([\rho];\rv_i)$ has been built, the sum over the $N$ electrons in the right-hand side of Eq.~\eqref{eq_esofr} is replaced by $N E_{\rm cell}([\rho];\rv)$.\cite{SeiPerKur-PRA-00} In the original derivation of the PC model \cite{SeiPerKur-PRA-00} this step was seen as a further approximation. We now know, thanks to the exact SCE formulation, that this is is not an approximation, but an exact feature of the $\lambda\to\infty$ limit, clarified in Eq.~\eqref{eq_changegauge}. Because of this transformation, the local electrostatic energy that the PC model is trying to approximate is then exactly the same as the one of the exchange-correlation hole of Eq.~\eqref{eq_endensSCE}.

It is important to stress that the PC cell {\em is not} an approximation to the exchange-correlation hole in the $\lambda\to\infty$ limit.\cite{SeiPerKur-PRA-00} However, we have now proved that its electrostatic energy (electron-background attraction plus background-background repulsion) {\em is} an approximation to the electrostatic potential of the exchange correlation hole, Eq.~\eqref{eq:defenergydensity}. This concept is further clarified in the Appendix, where the case of the uniform electron gas at low density is treated explicitly.

The simplest approximation to the PC cell is a sphere of uniform density $\rho(\rv)$ around the electron at position $\rv$ with a radius $r_s(\rv)=\left(\frac{4 \pi}{3}\rho(\rv)\right)^{-1/3}$ fixed by the condition that the fictitious positive background exactly neutralizes the electron at its center. This leads to the simple PC-LDA approximation\cite{SeiPerKur-PRA-00} 
\beq
\enerdens_{\rm PC}^{\rm LDA}(\rv)=-\frac{9}{10}\left(\frac{4\pi}{3}\right)^{1/3}\rho(\rv)^{1/3}.
\label{eq_PCLDA}
\eeq
If we approximate the dipole moment of the cell in terms of the gradient of the density and we set it equal to zero we obtain the PC-GGA expression\cite{SeiPerKur-PRA-00}
\beq
\enerdens_{\rm PC}^{\rm GGA}(\rv)=\enerdens_{\rm PC}^{\rm LDA}(\rv)+\frac{3}{350}\left(\frac{3}{4\pi}\right)^{1/3}\frac{|\nabla\rho(\rv)|^2}{\rho(\rv)^{7/3}}.
\label{eq_PCGGA}
\eeq

\begin{figure}
\includegraphics[width=5.cm,angle=-90]{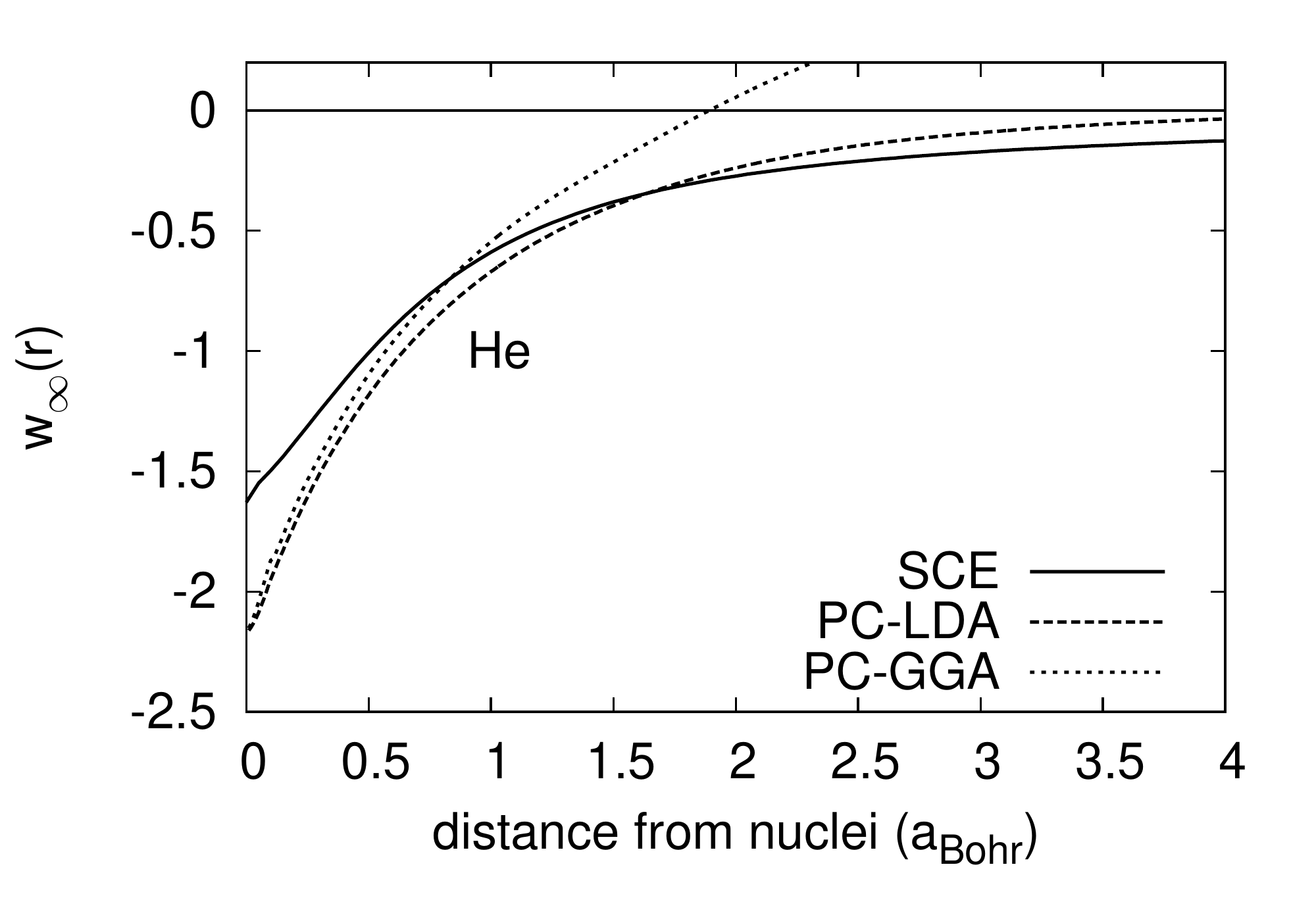}
\includegraphics[width=5.5cm,angle=-90,clip,trim= 2.8cm 0cm -.3cm 6cm]{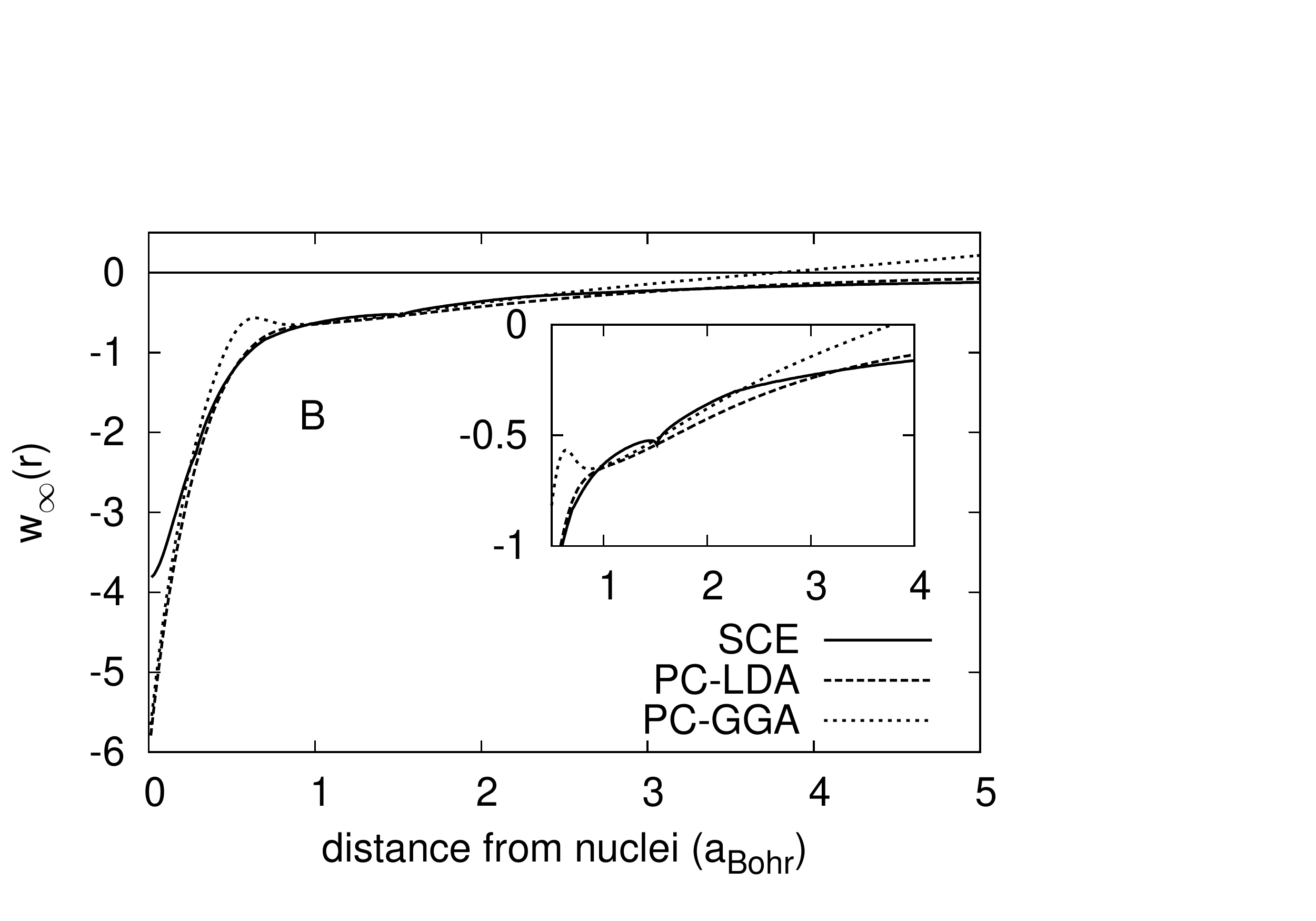}
\includegraphics[width=5.5cm,angle=-90,clip,trim= 2.8cm 0cm -.3cm 6cm]{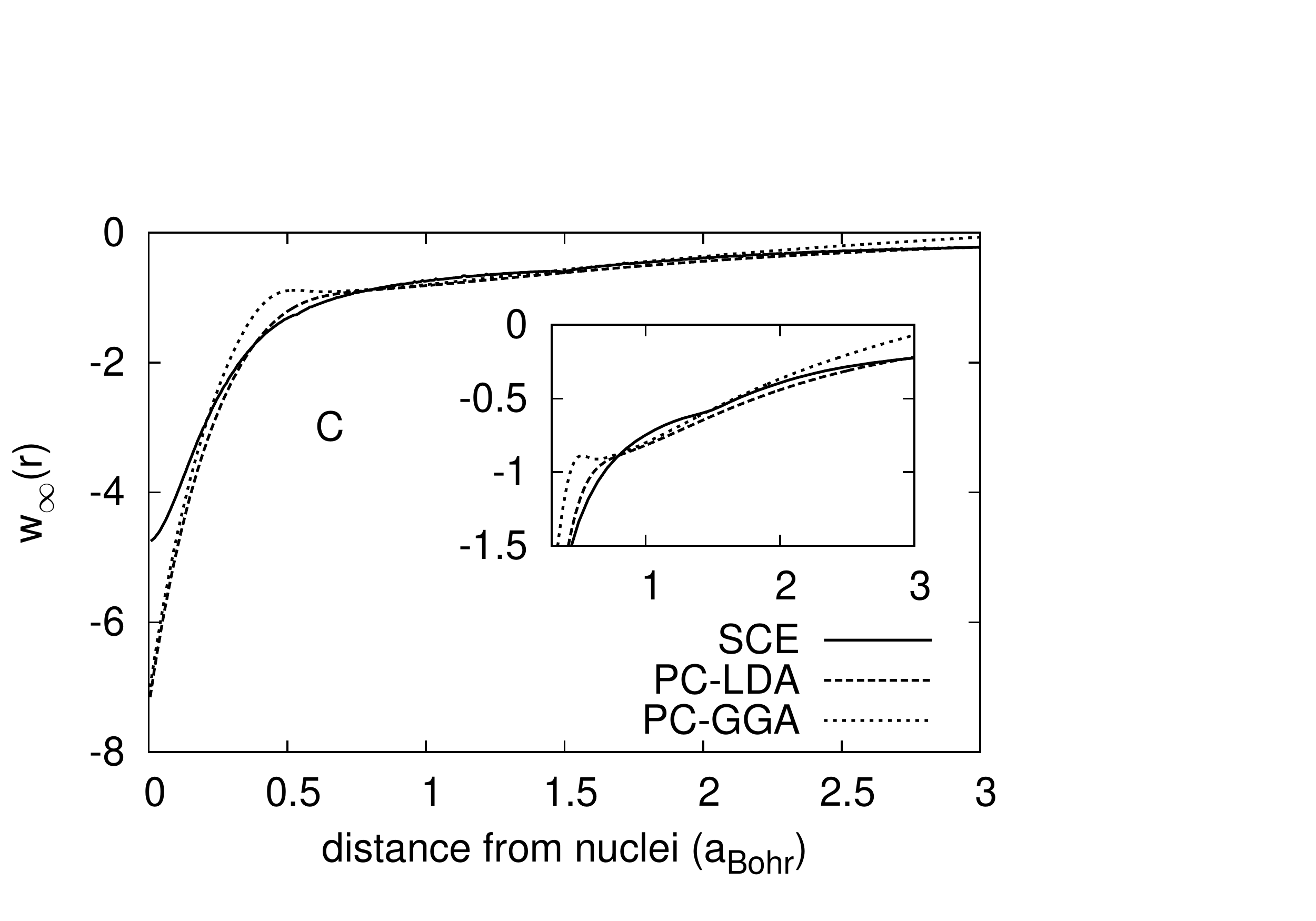}
\includegraphics[width=5.5cm,angle=-90,clip,trim= 2.8cm 0cm -.3cm 6cm]{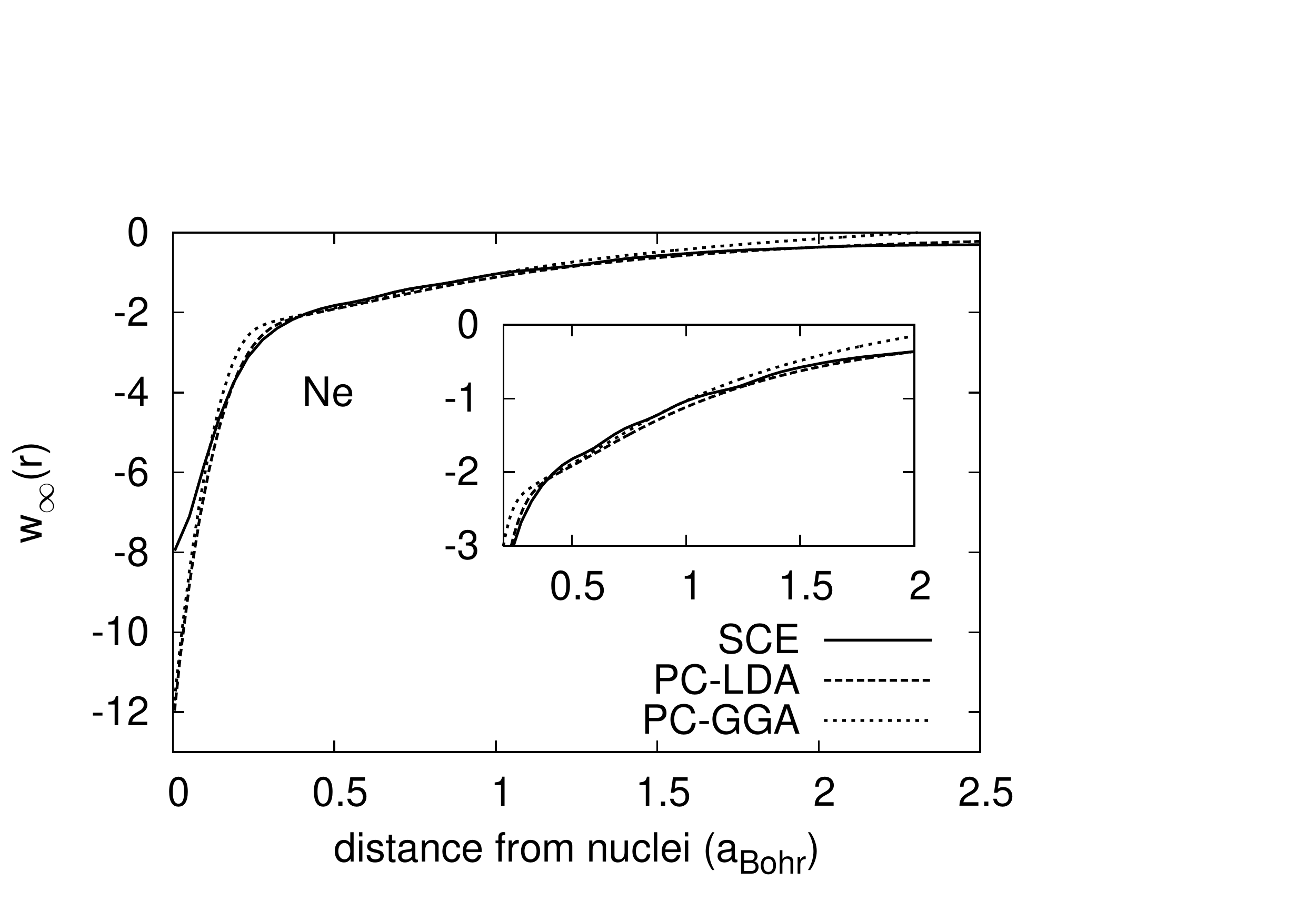}
\caption{Energy density in the gauge of the electrostatic potential of the exchange-correlation hole, Eq.~\eqref{eq:defenergydensity}, in the $\lambda\to\infty$ limit. The exact SCE result of Eq.~\eqref{eq_endensSCE} is compared with the PC-LDA and PC-GGA approximations of Eqs.~\eqref{eq_PCLDA}-\eqref{eq_PCGGA}.}
\label{fig_SCEandPC}
\end{figure}

In Fig.~\ref{fig_SCEandPC} we compare the exact $\lambda\to\infty$ energy densities of Eq.~\eqref{eq_endensSCE} with the PC-LDA and PC-GGA approximations of Eqs.~\eqref{eq_PCLDA}-\eqref{eq_PCGGA} for the He atom, the sphericalized B and C atoms and for the Ne atom, using accurate Hylleras and quantum Monte Carlo densities.\cite{FreHuxMor-PRA-84,AlsResUmr-PRA-98}\footnote{J. Toulouse, private communication. The sphericalized densities of the B and C atoms were obtained from variational
Monte Carlo using accurate optimized wavefunctions as described in Refs.~\onlinecite{TouUmr-JCP-07} and~\onlinecite{UmrTouFilSorHen-PRL-07}.}
We see that the PC model becomes a rather good approximation in the valence region of B, C and Ne, while being quite poor in the core region, and especially at the nucleus. The PC-LDA energy density is actually a better local approximation except close to the nucleus. The PC-GGA performs better globally (see Table \ref{tab_PCglobal}), but we clearly see that this is due to an error compensation between the core region and the intershell region.

The approximations done in the PC model are {\it i}) neglecting the cell-cell interaction, and {\it ii}) the gradient expansion of Eqs.~\eqref{eq_PCLDA}-\eqref{eq_PCGGA} which assumes a slowly varying density. At the nucleus, we can easily construct what would be the ``exact'' PC cell, so that we can at least remove approximation {\it ii}) and check the effect of approximation {\it i}) alone. The ``exact'' PC cell around the nucleus is the sphere $\Omega_1$ of radius $a_1$, with
\beq
\int_0^{a_1}4 \pi r^2\,\rho(r)\,d r=1,
\eeq
and the ``exact'' value of $\enerdens_{\rm PC}(r=0)$ is
\beq
\enerdens_{\rm PC}(r=0)=-\int_{\Omega_1}d\rv\frac{\rho(\rv)}{r}+\frac{1}{2}\int_{\Omega_1}d\rv\int_{\Omega_1}d{\bf r}'\frac{\rho(\rv)\rho(\rv')}{|\rv-\rv'|}.
\label{eq_wPC0}
\eeq
In Table~\ref{tab_PCr0} we compare the values at the nucleus from the exact SCE, the PC-LDA or PC-GGA (they become equal at the nucleus) and the result of Eq.~\eqref{eq_wPC0} for several atoms. We see that Eq.~\eqref{eq_wPC0} is very accurate for $N=2$ electrons: in this case, in fact, when the reference electron is at the nucleus, the other one is at infinity, so that the cell-cell interaction becomes indeed zero. For $N>2$, we see that the ``smearing hypothesis,'' i.e. the idea that the cell-cell interaction is negligeable, leads to some errors, although there is an improvement with respect to the gradient expansion of Eqs.~\eqref{eq_PCLDA}-\eqref{eq_PCGGA}, reducing the relative error of about a factor 2. Along these lines, one might try to construct an improved PC model that performs locally better than the PC-GGA, which, as said, achieves good global accuracy at the price of error compensation between different regions of space.
\begin{table}
\begin{tabular}{llll}
\hline \hline
& SCE & PC-LDA  & PC-GGA\\
\hline 
H$^-$  & $-0.569$ & $-0.664$    & $-0.559$\\
He & $-1.498$     & $-1.735$     & $-1.468$ \\
Li & $-2.596$     & $-2.983$    & $-2.556$\\
Be & $-4.021$     & $-4.561$    & $-3.961$\\ 
B &  $-5.706$     & $-6.412$    & $-5.650$ \\
C &  $-7.781$     & $-8.650$    & $-7.719$\\ 
Ne & $-19.993$     & $-21.647 $   & $-19.999$\\ 
\hline \hline
\end{tabular}
\caption{Global value $W_{\infty}[\rho]=\int\rho(\rv)\enerdens_{\infty}[\rho](\rv) d \rv$ for small atoms at different levels of approximation. The SCE corresponds to the exact value, Eq.~\eqref{eq_endensSCE}, while PC-LDA and PC-GGA correspond, respectively, to 
Eqs.~\eqref{eq_PCLDA} and \eqref{eq_PCGGA}.}
\label{tab_PCglobal}
\end{table}

\begin{table}
\begin{tabular}{llll}
\hline \hline
& $\enerdens_{\rm SCE}(r=0)$  & $\enerdens_{\rm PC}^{\rm GGA}(r=0)$  & $\enerdens_{\rm PC}(r=0)$\\
\hline 
H$^-$  & $-0.6825$ & $-0.9671$    & $-0.7157$\\
He & $-1.6883$     & $-2.1729$     & $-1.6672$ \\
Li & $-2.2041$     & $-3.4019$    & $-2.6396$\\
Be & $-3.1568$     & $-4.6578$    & $-3.6354$\\ 
B &  $-3.8230$     & $-5.8995$    & $-4.6190$ \\
C &  $-4.7727$     & $-7.1446$    & $-5.6050$\\ 
Ne & $-8.0276$     & $-12.119 $   & $-9.5463$\\ 
\hline \hline
\end{tabular}
\caption{Comparison of the values at the nucleus of the energy density in the gauge of the exchange-correlation hole potential in the strong-interaction limit for small atoms. The value $\enerdens_{\rm SCE}(r=0)$ corresponds to the exact expression  of Eq.~\eqref{eq_endensSCE}, the value $\enerdens_{\rm PC}^{\rm GGA}(r=0)$ is the PC gradient expansion approximation of Eqs.~\eqref{eq_PCLDA}-\eqref{eq_PCGGA} (the PC-LDA and PC-GGA are equal at the nucleus), and $\enerdens_{\rm PC}(r=0)$ is the value from the ``exact'' PC cell of Eq.~\eqref{eq_wPC0}.
}
\label{tab_PCr0}
\end{table}

\section{Energy densities along the adiabatic connection}
\label{sec_enedenslambda}

\subsection{Kohn-Sham ($\lambda=0$)}

At \textit{zero} coupling-strength the exact solution for the wave function $\Psi_0$ becomes a Slater determinant $\Phi=|\phi_1...\phi_N\rangle$, and the energy density $\enerdens_0[\rho](\rv)$ in the gauge of the exchange-correlation hole is given by the electrostatic potential of the KS exchange hole $h_{x}(\rv,\rv')$,
\begin{equation}
	\enerdens_0[\rho](\rv)=\frac{1}{2}\int \frac{h_{x}(\rv,\rv')}{|\rv-\rv'|}\,d\rv',
	\label{eq:energydensityKS}
\end{equation}
as the pair density simply writes
\begin{equation}
	P_2^0(\rv,\rv')=\rho(\rv)\rho(\rv')+\rho(\rv)h_{x}(\rv,\rv').
	\label{eq:PdensKS}
\end{equation}
One can use in Eq.~\eqref{eq:energydensityKS} the exact exchange hole built from a Hartee-Fock like expression in terms of the KS orbitals $\phi_i$, or a density functional approximation for $h_{x}(\rv,\rv')$, e.g., the one of Becke and Roussel.\cite{BecRou-PRA-89} These two choices would correspond, respectively, to construct a hyper-GGA and a meta-GGA functional from a local interpolation along the adiabatic connection.
 
The aim of the present work is a preliminary study of {\em exact} energy densities along the adiabatic connection. The exact KS orbitals and the corresponding non-interacting potential $\hat{V}^0_{\rm ext}$ for a given physical density can be found in an exact way, e.g,. by inversion of the KS equations\cite{WanPar-PRA-93,vanBae-PRA-94,GrivanBae-PRA-95,PeiNecWar-PRA-03,KadSto-PRA-04,AstSto-PRB-06} or by the use of Lieb's Legendre transform DFT formalism.\cite{Lie-IJQC-83,ColSav-JCP-99,TeaCorHel-JCP-10} 

For an ISI-like interpolation on the energy density, $\enerdens_0[\rho](\rv)$ will be a key ingredient. Additionally, knowledge of the next leading order in the asymptotic expansion of the local energy density around $\lambda=0$ is necessary, but not available yet. The next leading order in the asymptotic expansion constitutes an active field of research in our group (see also the discussion in Sec.~\ref{sec_conc}).

\subsection{Physical ($\lambda=1$)}

To compute the exact energy density at coupling-strength $\lambda=1$ we resort to
\begin{equation}
	\enerdens_1[\rho](\rv)=\frac{1}{2\rho(\rv)}\int \frac{P_2^1(\rv,\rv')}{|\rv-\rv'|}\,d\rv'-\frac{1}{2}\int \frac{\rho(\rv')}{|\rv-\rv'|}\,d\rv',
	\label{eq:energydensityphys}
\end{equation}
with the pair density given by the full many-body wave function $\Psi_1$ in Eq. \eqref{eq_P2}. The density $\rho_1(\rv)$ out of $P_2^1(\rv,\rv')$ defines the density $\rho(\rv)=\rho_1(\rv)$ to be held constant along the adiabatic connection.

The exact $\enerdens_1[\rho](\rv)$ can serve as benchmark for models on $\enerdens_\lambda[\rho](\rv)$ but also gives an estimate of the importance of the strong-interaction limit in the $\enerdens_\lambda[\rho](\rv)$ model. If the physical system is close to the KS one, correlation is less important and already Hartree-Fock should perform well. In this case, we expect that inclusion of the $\lambda\rightarrow\infty$ information in the $\enerdens_\lambda[\rho](\rv)$ model does not lead to a major improvement. In contrast, for stronger correlated system the physical energy density should tend more towards the $\lambda\rightarrow\infty$ limit and the SCE concept can provide useful input for an accurate model for $\enerdens_\lambda[\rho](\rv)$. The relevance of the strong-interaction limit will be discussed in the next section.


\subsection{Results}

\subsubsection{Coulomb external potential}

We have performed full-CI calculations in an aug-cc-pVTZ basis for some two and four electron atoms within the Gamess-US package\cite{GAMESS} to obtain an accurate ground state wave function for the physical interaction strength. Starting from this, we are able to calculate the energy density in the gauge of the exchange-correlation hole for $\lambda=0,1,\infty$.

At $\lambda=1$ we calculate the energy density from the full-CI pair density, Eq.~\eqref{eq:energydensityphys}, by a program similar to the one used for the calculation of $v_{\rm cond}$ in Ref.~\onlinecite{BuiBaeSni-PRA-89}.

For the energy density at $\lambda=0$, Eq.~\eqref{eq:energydensityKS}, we have to compute the single particle KS orbitals corresponding to the full-CI density first. In the case of 2 electron atoms they are readily constructed by the simple relation
\beq
	\phi(\rv)=\sqrt{\dfrac{\rho(\rv)}{2}}.
\eeq
For the four electron atoms we choose the scheme of van~Leeuwen, Baerends and Gritsenko\cite{vanBae-PRA-94,GrivanBae-PRA-95} to invert the KS equations.
In the strong-interaction limit we calculate the energy density within the SCE concept, see Sec.~\ref{sec:SCEexact} and Refs.~\onlinecite{SeiGorSav-PRA-07,GorVigSei-JCTC-09}.

\begin{figure}
	\begin{center}
	\includegraphics[width=5.cm,angle=-90]{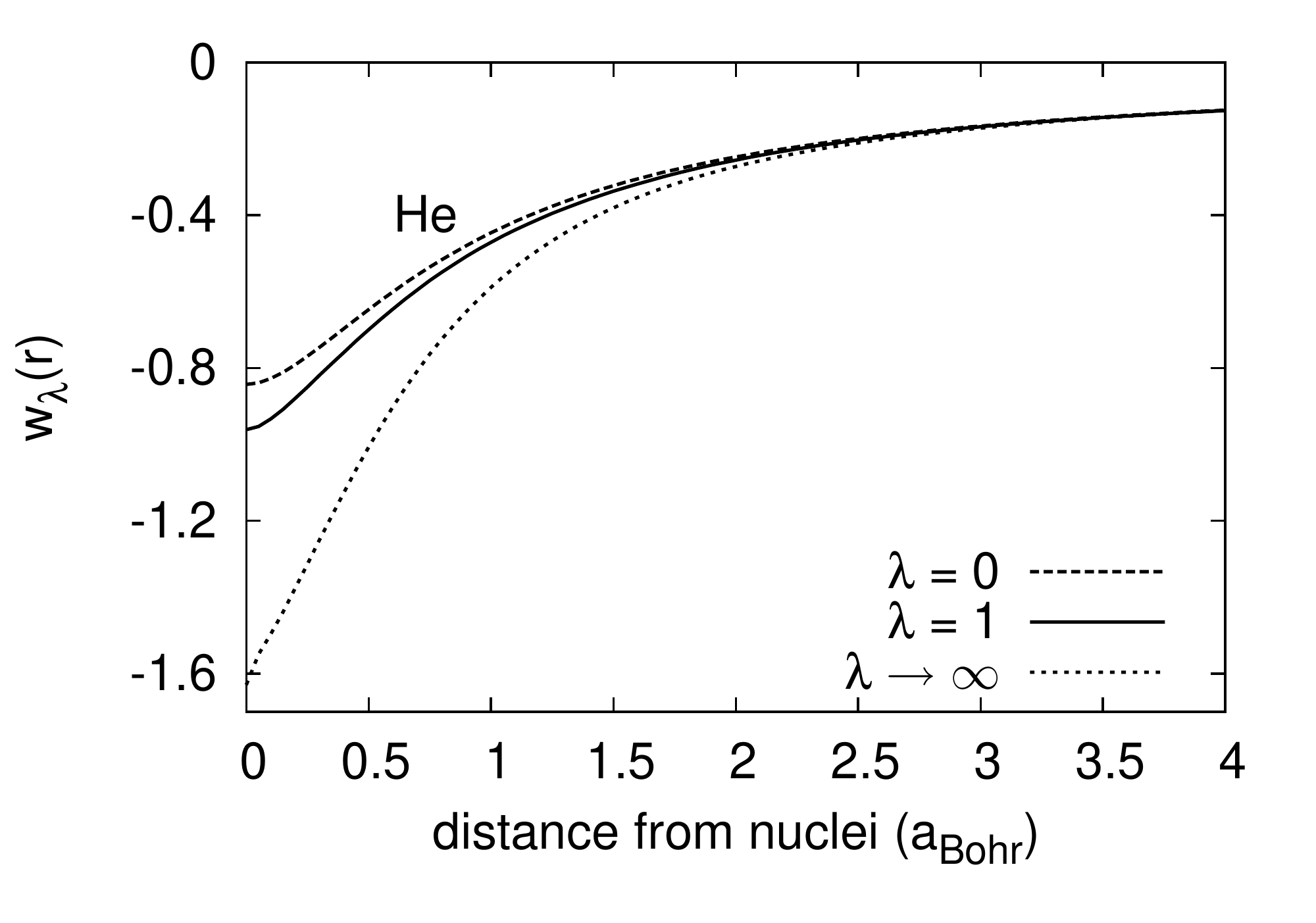}
	\includegraphics[width=5.cm,angle=-90]{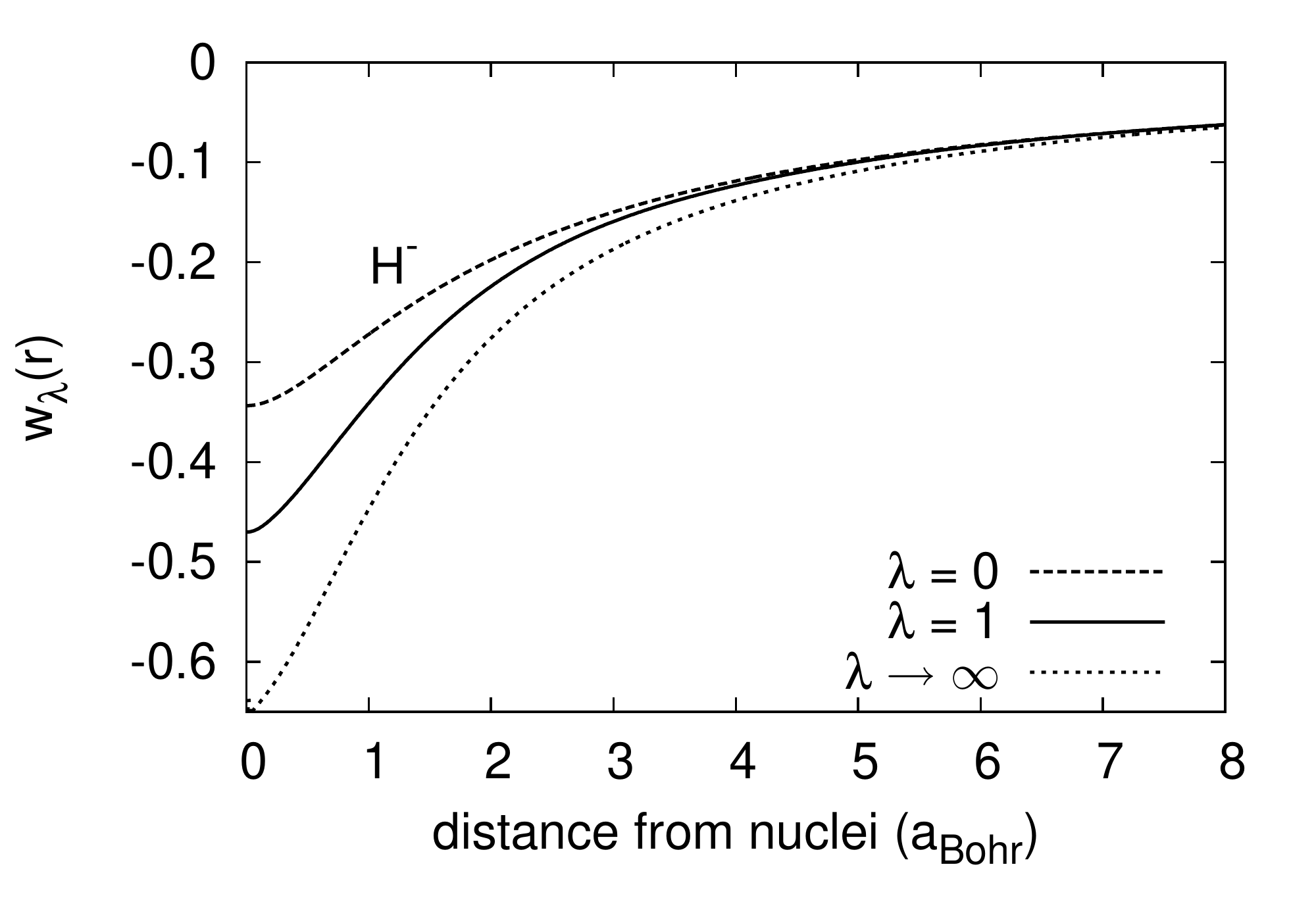}
    \includegraphics[width=5.5cm,angle=-90,clip,trim= 2.8cm 0cm -.3cm 6cm]{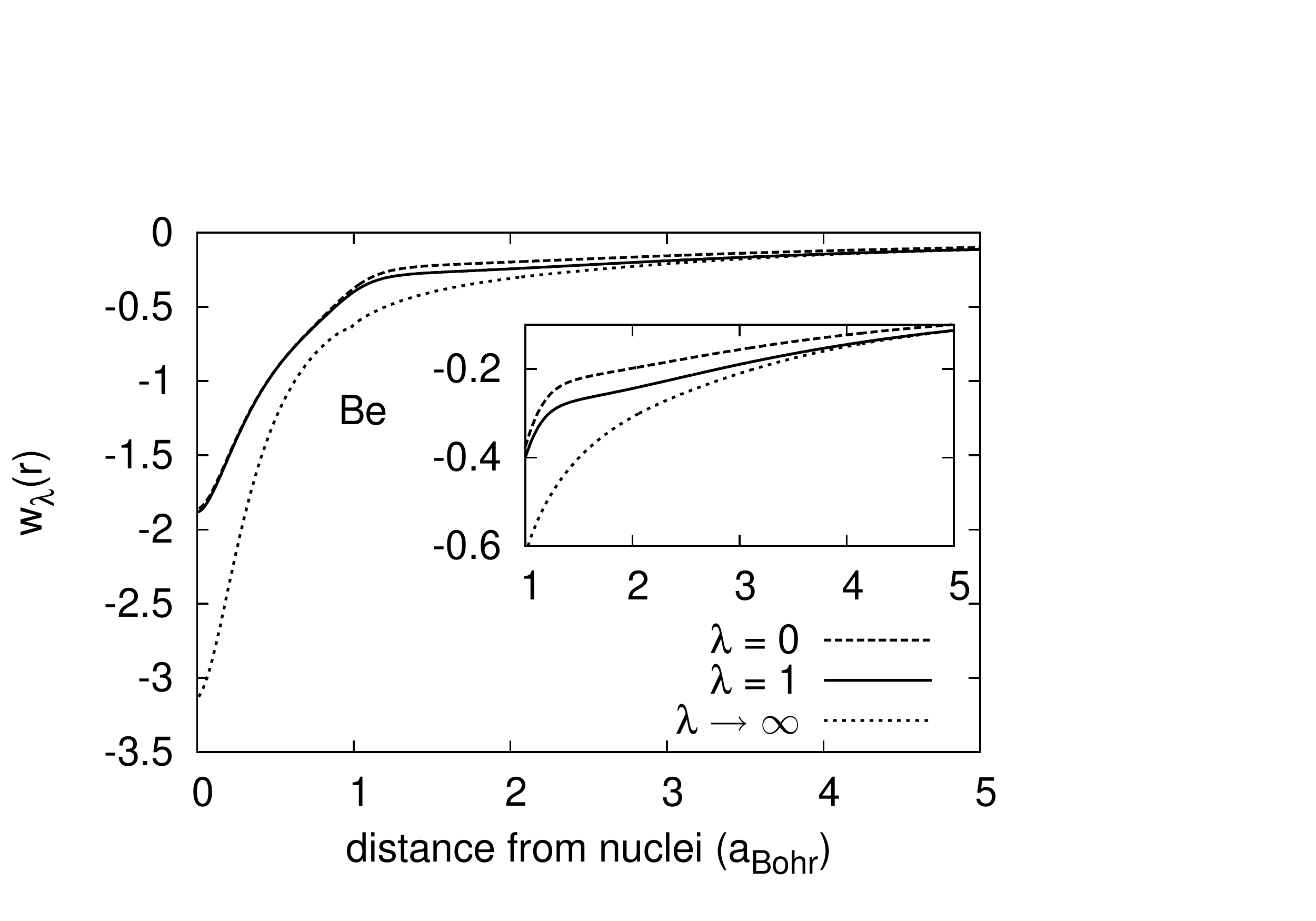}
	\includegraphics[width=5.5cm,angle=-90,clip,trim= 2.8cm 0cm -.3cm 6cm]{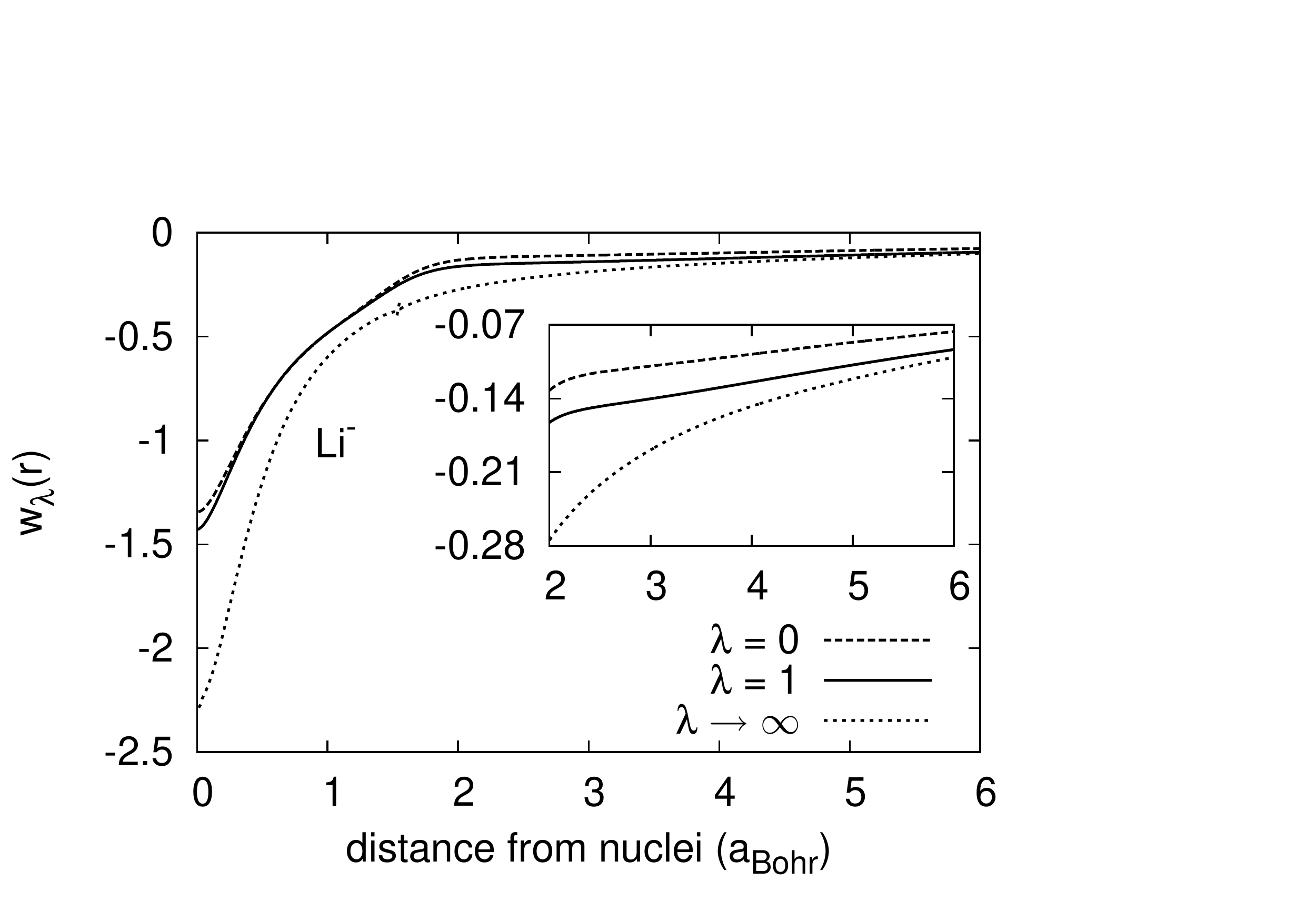}
   	\caption{Energy densities in the gauge of the electrostatic potential of the exchange-correlation hole for accurate full-CI densities (aug-cc-pVTZ) and coupling strength $\lambda=0,1$ and $\infty$.}
	\label{fig_l=0,1,inf}
	\end{center}
\end{figure}

In Fig.~\ref{fig_l=0,1,inf} we show the energy densities for $\lambda=0,1,\infty$ for two- and four-electron atoms. 
As expected, He and Be are relatively weakly correlated, and their $\lambda=1$ energy densities are much closer to the KS ones than to the SCE ones. Here a description at the Hartree-Fock level is very reasonable and gives indeed at least $98.5\%$ of the total energy. The anion H$^-$, instead, being a system with a more diffuse density and thus more correlated, has a physical energy density that is much more in between the KS and the SCE curves, with an Hartree-Fock treatment giving only $94\%$ of the total energy. Here we expect the inclusion of the information from the strong-interaction limit to be important. The valence regions of Be and Li$^-$ (see the insets in Fig.~\ref{fig_l=0,1,inf}) can also be better described by a proper inclusion of the $\lambda=\infty$ information.

\subsubsection{Harmonic external potential}

\begin{figure}
	\begin{center}
	\includegraphics[width=5.5cm,angle=-90,clip,trim= 2.8cm 0cm -.3cm 6cm]{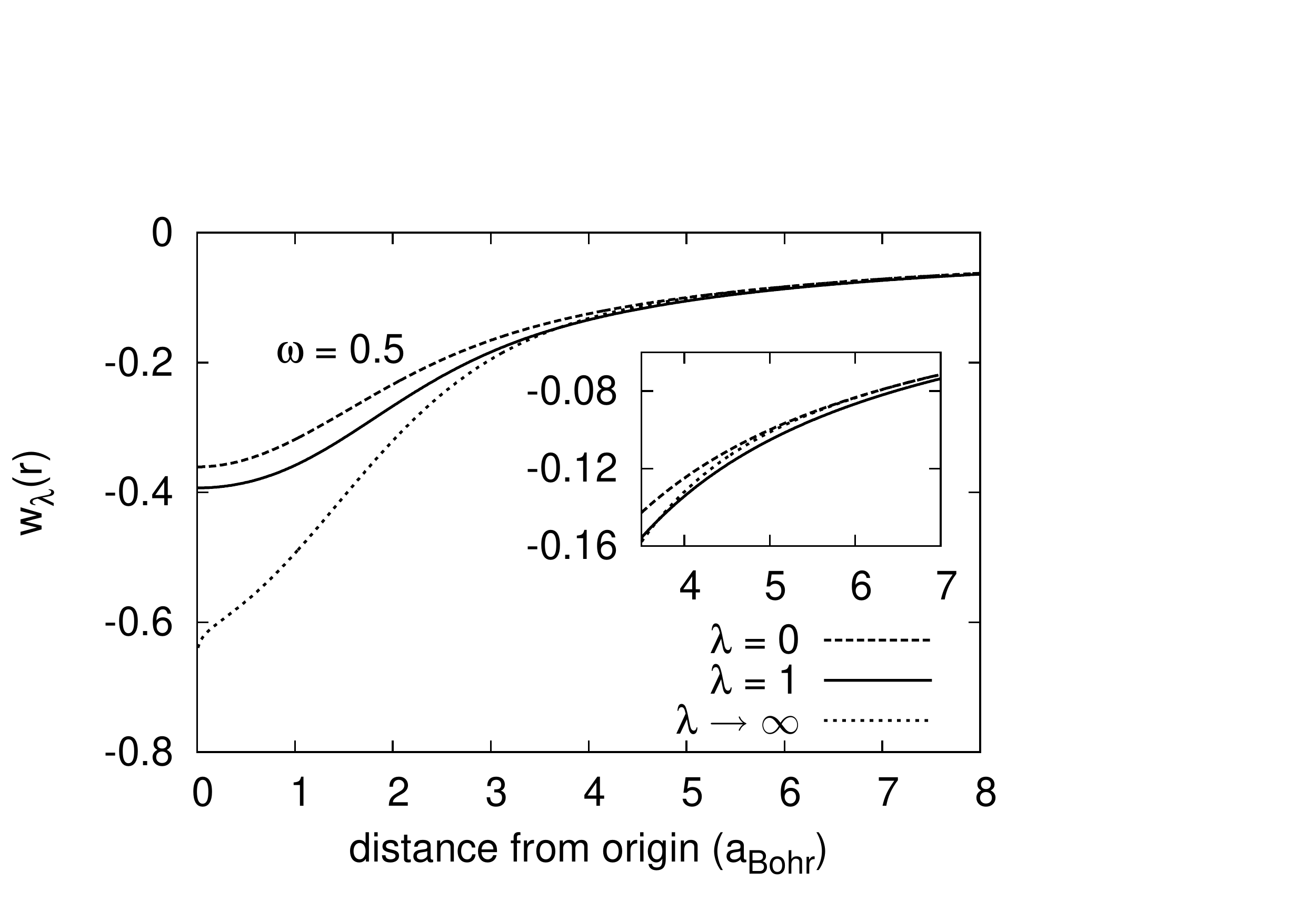}
	\includegraphics[width=5.5cm,angle=-90]{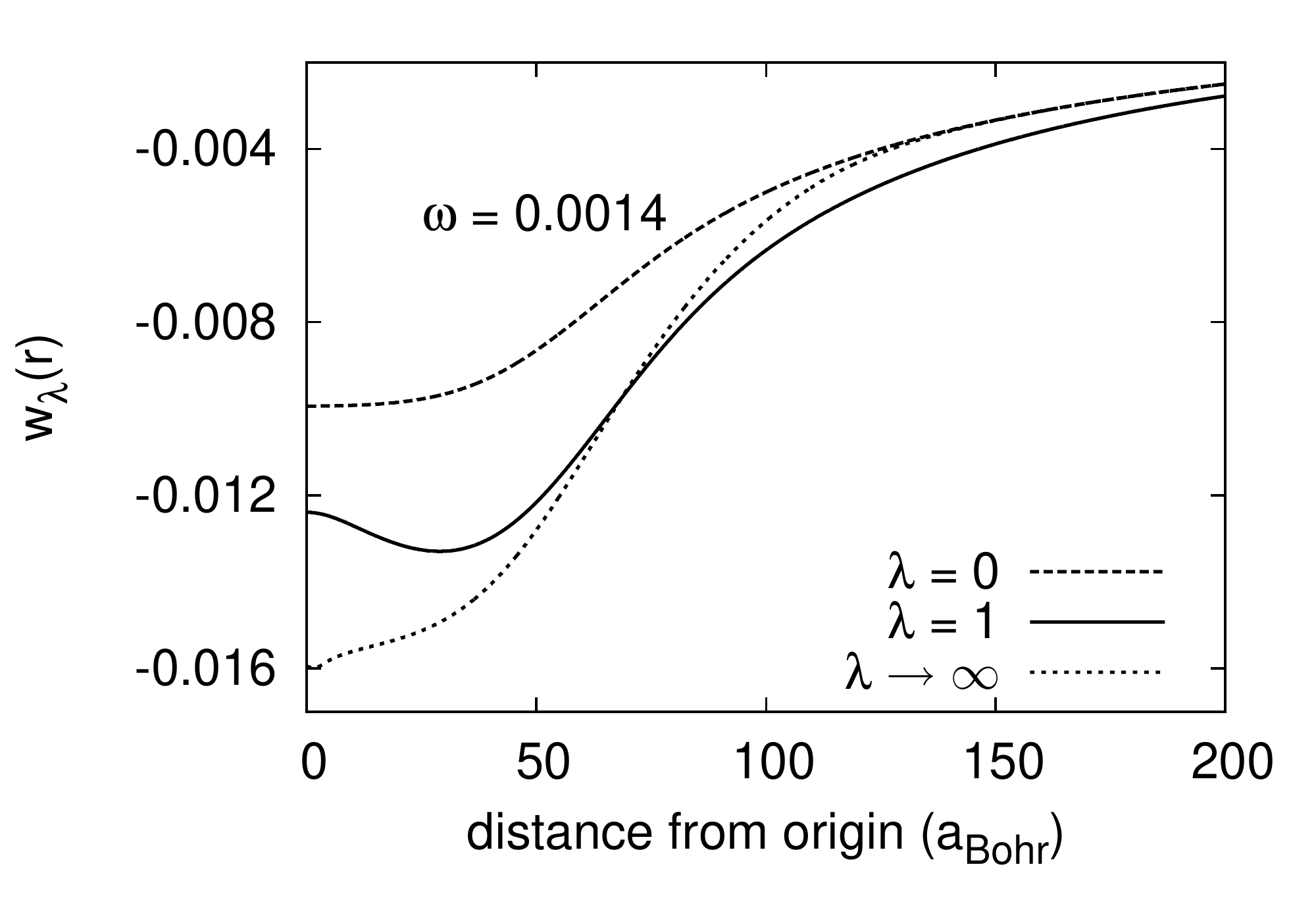}
	\caption{Energy densities in the gauge of the electrostatic potential of the exchange-correlation hole for a quantum dot with $N=2$ electrons with less pronounced correlation ($\omega=0.5$) and pronounced correlation ($\omega=0.0014$).}
	\label{fig:hooke}
	\end{center}
\end{figure}

Another useful class of systems to investigate the impact of the strong-interaction limit on the physical energy density is given by model quantum dots, where electrons are confined by a harmonic potential and correlation gains importance as the spring constant is lowered. We have computed the energy density for $N=2$ electrons in three dimensions for spring constants for which an analytic solution for the wave function can be found.\cite{Tau-PRA-93} The results are displayed in Fig.~\ref{fig:hooke} for the largest and smallest spring constant considered.
As expected, the physical energy density comes closer to the SCE energy density in the stronger correlated case. Additionally, a remarkable feature we observe is that the physical energy density crosses the SCE energy density. By intuition one would expect the physical energy density to be always in between the KS and SCE energy densities, as the KS energy density represents the weakest possible correlation and the SCE energy density the strongest possible correlation in the given density. However, the wave functions are chosen according to the global quantities
\begin{eqnarray}
	\min_{\Psi\rightarrow\rho}\langle\Psi|\hat{T}|\Psi\rangle & \Rightarrow & \Psi_{\rm KS} \\
	\min_{\Psi\rightarrow\rho}\langle\Psi|\hat{V}_{ee}|\Psi\rangle & \Rightarrow & \Psi_{\rm SCE} 
	\label{eq_VeeSCEconstr}\\
	\min_{\Psi\rightarrow\rho}\langle\Psi|\hat{T}+\hat{V}_{ee}|\Psi\rangle & \Rightarrow & \Psi_{\lambda=1},
\end{eqnarray}
yielding the global inequalities
\beq
\langle\Psi_{\rm SCE}|\hat{V}_{ee}|\Psi_{\rm SCE}\rangle\leq\langle\Psi_{\lambda=1}|\hat{V}_{ee}|\Psi_{\lambda=1}\rangle\leq	\langle\Psi_{\rm KS}|\hat{V}_{ee}|\Psi_{\rm KS}\rangle.
\eeq	
Locally, these inequalities can be violated without violating the global ones, and hence the physical energy density can go below the SCE energy density. 

\begin{figure}
	\begin{center}
	\includegraphics[width=5.5cm,angle=-90,]{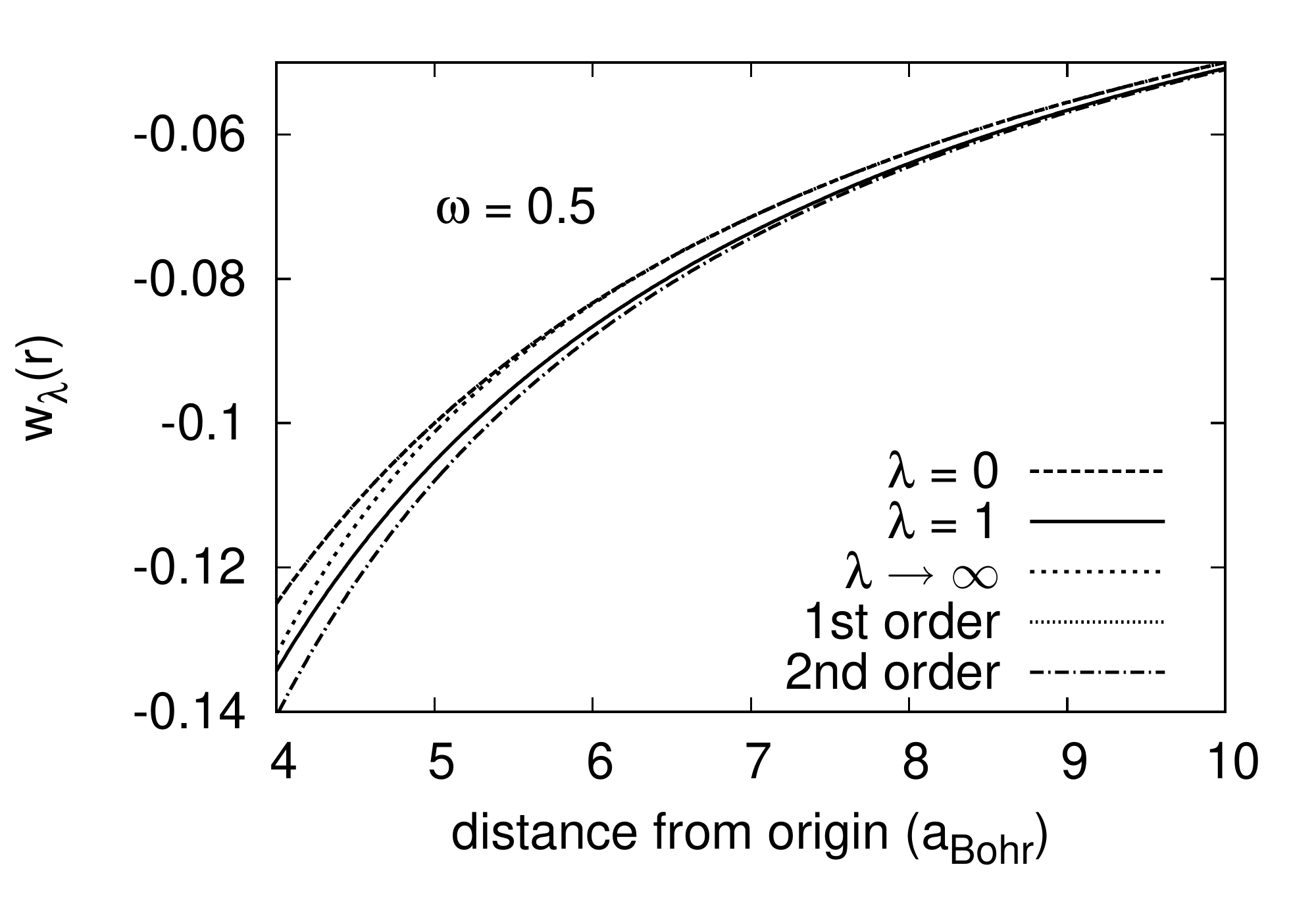}
	\caption{Energy density in the gauge of the electrostatic potential of the exchange-correlation hole for a $N=2$-electron model quantum dot calculated with the full KS, physical and SCE pair density and with the asymptotic expansion of the physical pair density, Eq.~\eqref{eq:Passympt}. The KS and 1st order curve lie on top of each other.}
	\label{fig:hooke_asympt}
	\end{center}
\end{figure}

The crossing feature can be attributed to polarization effects, that are present in the physical case, but are not contained in the KS and SCE pair density. In the KS case we rely on the independent-particle picture and polarization of a particle due to other particles is not described by this model. Whereas for the SCE, we build on point particles that are a priori unpolarizable. To underline this argument we have computed the asymptotic behavior of the physical energy density for the quantum dot with $\omega=0.5$ by use of the asymptotic expansion of the physical pair density\cite{ErnBurPer-JCP-96}
\beq
	\dfrac{P(\rv,\rv')}{\rho(\rv)\rho_{N-1}(\rv')}\rightarrow1-2\dfrac{r'}{r}\cos(\Delta\Omega)+\cdots\quad(r\rightarrow\infty),
	\label{eq:Passympt}
\eeq
where $\rho_{N-1}(\rv')$ is the density of the $N-1$-particle system and $\Delta\Omega$ the angle between $\rv$ and $\rv'$. The second order term in Eq.~\eqref{eq:Passympt} represents the polarization correction. As can be seen from Fig.~\ref{fig:hooke_asympt}, for large distances $|\rv-\rv'|$ the energy density in the KS and SCE case behaves like $\nicefrac{-1}{r}$, which corresponds to the first order in the expansion of the physical pair density. Inclusion of the second order term in the energy density gives essentially the physical behavior and deviates from the KS and SCE energy densities. 

Although the crossing happens in a region in which the density is very small and thus with an almost negligible energetic contribution, the analysis presented here can be helpful in constructing models for $\enerdens_\lambda[\rho](\rv)$. Notice that, instead, with the Coulomb external potential we always observed, so far, the expected behavior $\enerdens_{\lambda\to\infty}(\rv)\le \enerdens_{\lambda=1}(\rv)\le \enerdens_{\lambda=0}(\rv)$ everywhere.

\section{The local form of the Lieb-Oxford bound}
\label{sec_LO}
The Lieb-Oxford (LO) bound\cite{Lie-PLA-79,LieOxf-IJQC-81,ChaHan-PRA-99} is a rigorous lower bound to the indirect part of the electron-electron repulsion energy $\tilde{W}[\Psi]$ associated with a given many-electron wave function $\Psi$, 
\beq
\tilde{W}[\Psi]\equiv\langle\Psi|\hat{V}_{ee}|\Psi\rangle-U[\rho_\Psi]\ge -C \int d\rv\,\rho_\Psi(\rv)^{4/3},
\eeq
where $\rho_\Psi(\rv)$ is the density obtained from the wave function $\Psi$. The positive constant $C$ is rigorously known to have a value\cite{LieOxf-IJQC-81,ChaHan-PRA-99} $C\le 1.679$. It has been suggested\cite{Per-INC-91,RasPitCapPro-PRL-09} that a tighter bound can be obtained by taking the value of $C$ that corresponds to the low-density limit of the uniform electron gas, $C\approx1.44$, since the bound is known to be more challenged when the number of electrons increases\cite{LieOxf-IJQC-81} and when the system has low density.\cite{LevPer-PRB-93}

The LO bound translates into a lower bound for the exchange and exchange-correlation functionals,\cite{Per-INC-91,LevPer-PRB-93}
\beq
E_x[\rho]\ge E_{xc}[\rho]\ge -C \int d\rv\,\rho(\rv)^{4/3},
\label{eq_LOExc}
\eeq
simply because $E_x[\rho]=W_{\lambda=0}[\rho]$ is the indirect Coulomb repulsion of the Slater determinant of KS orbitals, and $E_{xc}[\rho]$ is the sum of the indirect Coulomb repulsion of the physical wave function, $W_{\lambda=1}[\rho]$, plus the correlation correction to the kinetic energy which is always positive.

The way the LO bound is used in the construction of approximate functionals is, usually (with the exception of Ref.~\onlinecite{OdaCap-PRA-09}), by imposing it locally (see, e.g., Refs.~\onlinecite{PerBurErn-PRL-96,HauOdaScuPerCap-JCP-12}). That is, a given approximate exchange-correlation functional, $E_{x(c)}^{\rm DFA}[\rho]=\int \rho(\rv)\,\epsilon_{x(c)}^{\rm DFA}(\rv)\,d\rv$, is required to satisfy
\beq
\epsilon_{x(c)}^{\rm DFA}(\rv)\ge -C \,\rho(\rv)^{1/3}.
\label{eq_localLO}
\eeq
This is a sufficient condition to ensure the global bound of Eq.~\eqref{eq_LOExc}, but it is by no means necessary (see, e.g., Ref.~\onlinecite{ZhaYan-PRL-98}). In other words, there is no proof that a local version of the LO bound should hold. Actually, before even asking wether a local version of the LO bound should hold or not, we need to understand to which definition (gauge) of the energy density should apply the local LO bound of Eq.~\eqref{eq_localLO}. In fact, since energy densities are not uniquely defined, the inequality \eqref{eq_localLO} should be satisfied only for a well defined gauge: one can indeed always add to $\epsilon_{x(c)}^{\rm DFA}(\rv)$ a quantity that integrates to zero and violates Eq.~\eqref{eq_localLO} in some region of space.

We argue here that {\em i}) the gauge of the local LO bound is the conventional one of the electrostatic energy of the exchange-correlation hole, and {\em ii}) that the local LO bound is certainly violated, at least in the tail region of an atom or of a molecule, and in the bonding region of a stretched molecule. The argument behind point {\em i}) is the following. For a given density $\rho$, the wave function $\Psi[\rho]$ that maximally challenges\cite{RasSeiGor-PRB-11} the LO bound is the one that minimizes the expectation $\langle\Psi[\rho]|\hat{V}_{ee}|\Psi[\rho]\rangle$, i.e., by definition,  $\Psi_{\rm SCE}[\rho]$. In fact, we also have
\beq
E_x[\rho]\ge E_{xc}[\rho]\ge W_\infty[\rho]\ge -C \int d\rv\,\rho(\rv)^{4/3}.
\eeq
In Sec.~\ref{sec_strong} we have discussed the energy density associated to $W_\infty[\rho]$ in the gauge of the electrostatic potential of the exchange-correlation hole. We have also shown that this energy density can be approximated by the PC model that considers the electrostatic energy of a cell around the reference electron of positive charge $\rho_+(\rv)=\rho(\rv)$. The LDA version of this approximation has exactly the same form of the local LO bound. Moreover, the recently suggested value\cite{RasPitCapPro-PRL-09} $C\approx 1.44$ is extremely close to the one of the PC-LDA model, $C_{\rm PC}\approx1.45$. Notice that the fact that the PC model is in the gauge of the electrostatic energy of the exchange-correlation hole follows from the properties of the strong-interaction limit of DFT, in particular Eq.~\eqref{eq_changegauge}. If the PC model is an approximation in this gauge, and if the LO bound is locally equal to it, then conclusion {\em i}) should follow. 

\begin{figure}
	\begin{center}
	\includegraphics[width=5.5cm,angle=-90]{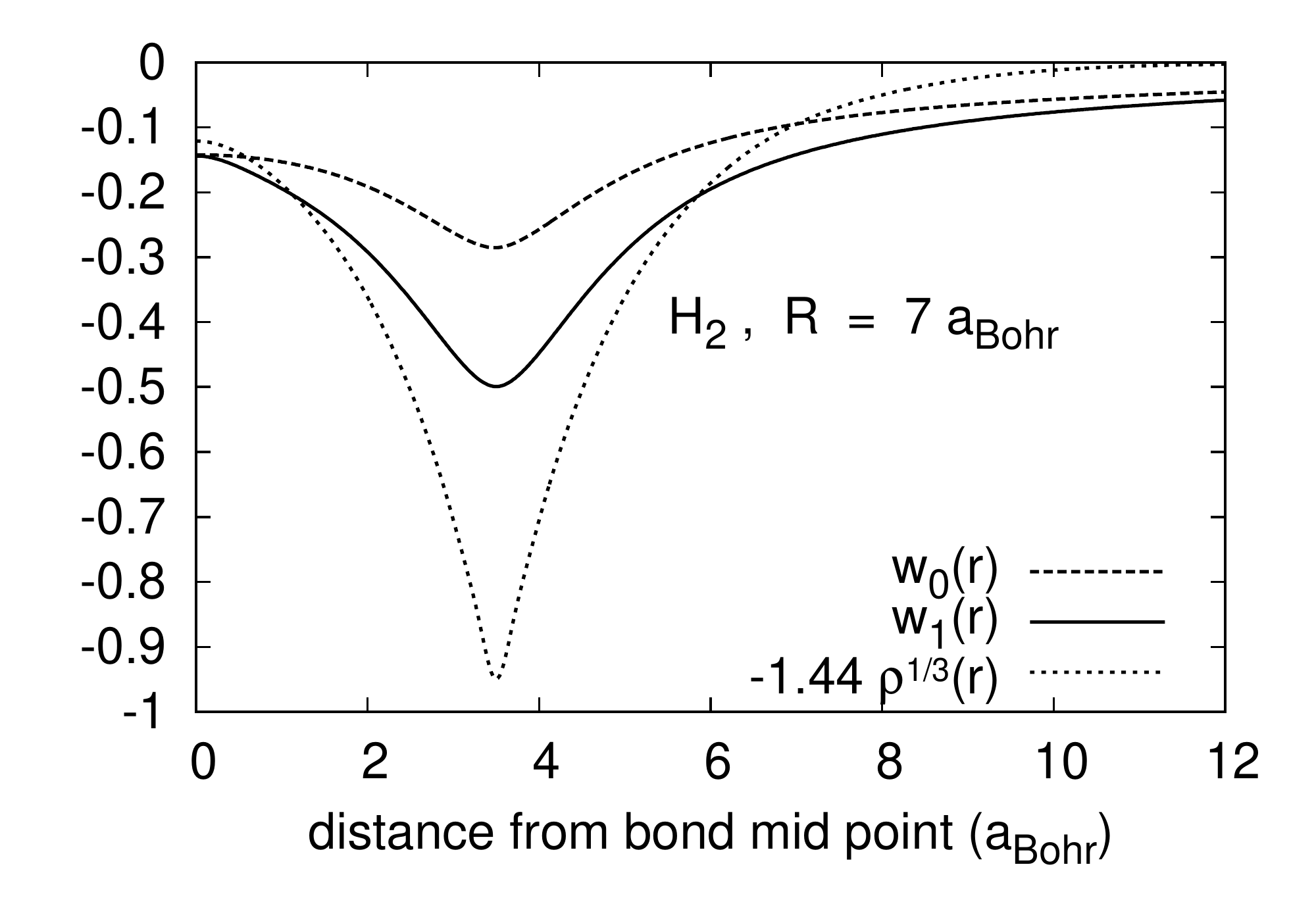}
	\includegraphics[width=5.5cm,angle=-90]{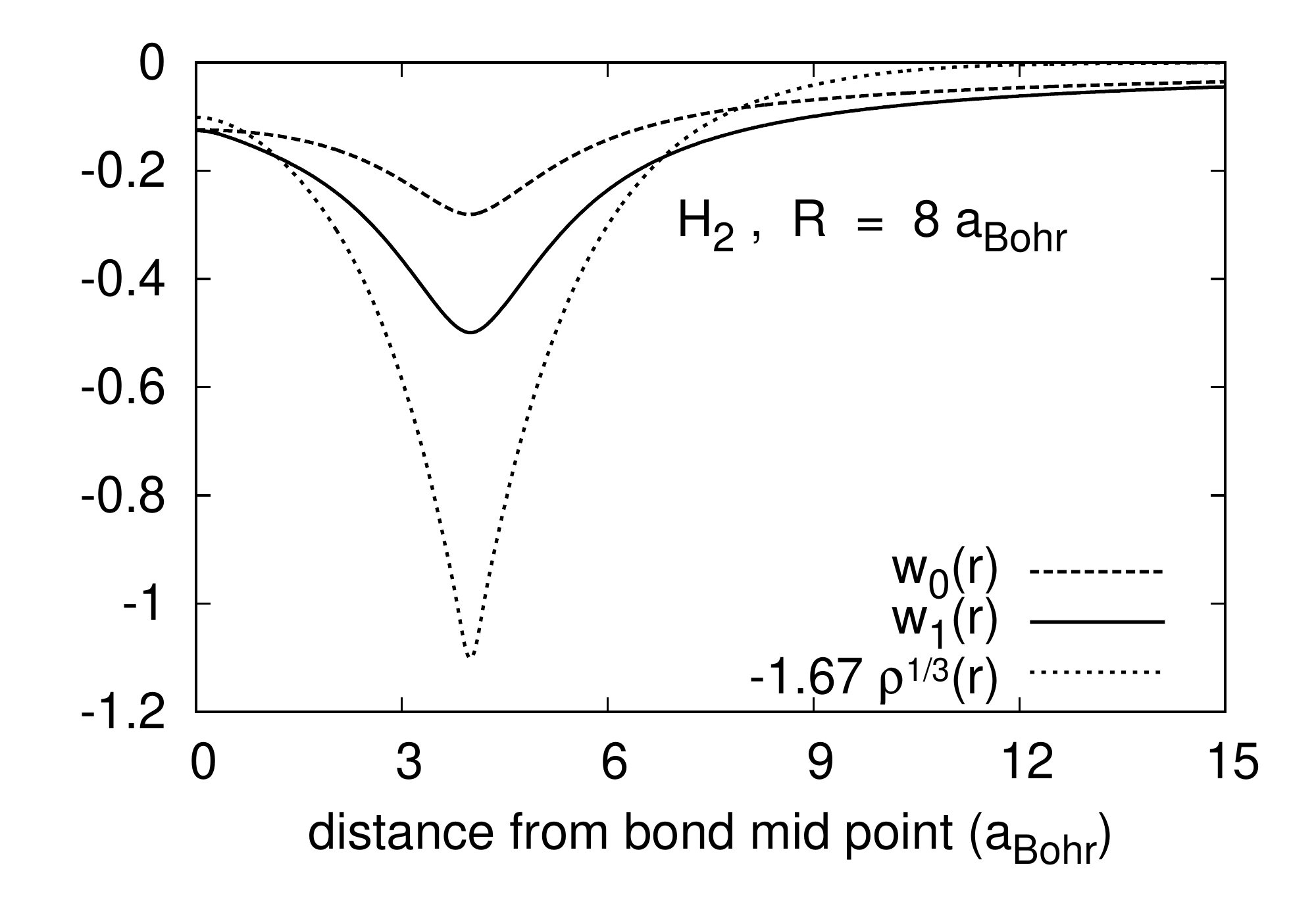}
\caption{Violation of the local form of the Lieb-Oxford bound for the stretched H$_2$ molecule.}
	\label{fig_h2LO}
	\end{center}
\end{figure}

We then easily see that the local LO bound of Eq.~\eqref{eq_localLO} is certainly violated in the tail region of an atom or a molecule, where the exact energy density in the conventional exchange-correlation hole gauge goes like $-1/r$ while the right-hand side of Eq.~\eqref{eq_localLO} decays exponentially. The local bound is also violated in the bond region of a stretched molecule. As an example, we show in Fig.~\ref{fig_h2LO} the energy densities of the stretched H$_2$ molecule for $\lambda=0$ and $\lambda=1$: with $C=1.44$ the local bound is violated in the bond region when the internuclear distance is $R\gtrsim 7$~a.u., and with $C=1.67$ when $R\gtrsim 8$~a.u.

As a concluding remark, we can say that it is very difficult, or maybe even impossible, to find a rigorous local lower bound for the energy density. In fact, we have just seen in Sec.~\ref{sec_enedenslambda} that, at least for the harmonic external potential, it is not even true that $\enerdens_{\lambda=1}(\rv)\ge \enerdens_{\infty}(\rv)$ everywhere.
This means that even if we maximize the correlation between the electrons we do not construct a rigorous local lower bound, but only a global one.

\section{Conclusions and Perspectives}
\label{sec_conc}
We have derived an exact expression for the energy density in the strong-interaction limit of DFT in the gauge of the exchange-correlation-hole electrostatic potential, and we have computed it for small atoms and model quantum dots. A careful analysis of the point-charge plus continuum (PC) model showed that this approximation is formulated in the same gauge, and a comparison with the exact results showed that it is locally reasonable.

Our formalism also allowed to analyze the local version of the Lieb-Oxford bound: we were able to assess to which gauge should the local LO bound correspond, then showing that it is certainly violated. Our findings are in agreement with (and give formal support to) the very recent results of Vilhena {\it et al.}\cite{VilRasLehMar-arxiv-12} (which only appeared when this manuscript was completed). More generally,  our results suggest that it is very difficult (if not impossible) to derive a rigorous local lower bound for the energy density.
  
We have also discussed the idea of a local interpolation along the adiabatic connection. The values of the local energy density in the same gauge at $\lambda=0$ and $\lambda=\infty$ are now available, either exactly or in an approximate way. Even if we have found that in the harmonic external potential the physical energy density is not always in between the $\lambda=0$ and the $\lambda=\infty$ curves, the regions of space in which the expected order is reversed are energetically not important. In the external Coulomb potential we have found, instead, the expected behavior $\enerdens_{\lambda\to\infty}(\rv)\le \enerdens_{\lambda=1}(\rv)\le \enerdens_{\lambda=0}(\rv)$ everywhere.

To really be able to build a local interpolation, at least the slope at $\lambda=0$, and possibly the next leading term at $\lambda=\infty$, are also needed in a local form, and in the same gauge. A first step towards the construction of a local slope at $\lambda=0$ is to produce exact results for this quantity, crucial to assess approximations. This can be achieved with the Legendre transform techniques developed in Refs.~\onlinecite{TeaCorHel-JCP-09,TeaCorHel-JCP-10} and it is the object of a current project. A possible way, then, to construct an approximate local slope is to use the so-called ``extended Overhauser model''\cite{GorPer-PRB-01,DavPolAsgTos-PRB-02,GorSav-PRA-05} locally, in a perturbative way. A local next leading term at $\lambda=\infty$ can also be constructed by deriving the exact exchange-correlation hole corresponding to the wave function of the zero-point oscillations, discussed in Ref.~\onlinecite{GorVigSei-JCTC-09}. All this will be investigated in future work.

\section*{Acknowledgments}

The authors thank Robert van Leeuwen for sharing the program for the inversion of the KS equations. We are grateful to Oleg Gritsenko for helpful assistance in developing the program for the calculation of the physical energy density. AM acknowledges Evert Jan Baerends for his warm hospitality at the Postech University in Pohang, Korea, where part of this work was done. PG-G thanks Gaetano Senatore for useful discussions. This work was supported by the Netherlands Organization for Scientific Research (NWO) through a Vidi grant.

\appendix

\section{PC cell and xc-hole}
In this Appendix we clarify the difference between the exchange-correlation hole and the PC cell by considering the uniform electron gas in the extreme low-density limit, further extending the argument already given in the Appendix of Ref.~\onlinecite{SeiPerKur-PRA-00}.

More than seventy years ago, Wigner\cite{Wig-PR-34,Wig-TFS-38} pointed out that electrons embedded in a compensating uniformly charged background would crystallize at sufficiently low values of the density $\rho$. The SCE construction can be seen as nothing else than the Wigner idea generalized to a nonuniform density $\rho(\rv)$. Indeed, in Ref.~\onlinecite{GorVigSei-JCTC-09} the SCE formalism is presented as a ``floating'' Wigner crystal in a non-euclidean space, with the metric determined by the density $\rho(\rv)$.
\begin{figure}
\includegraphics[width=8cm]{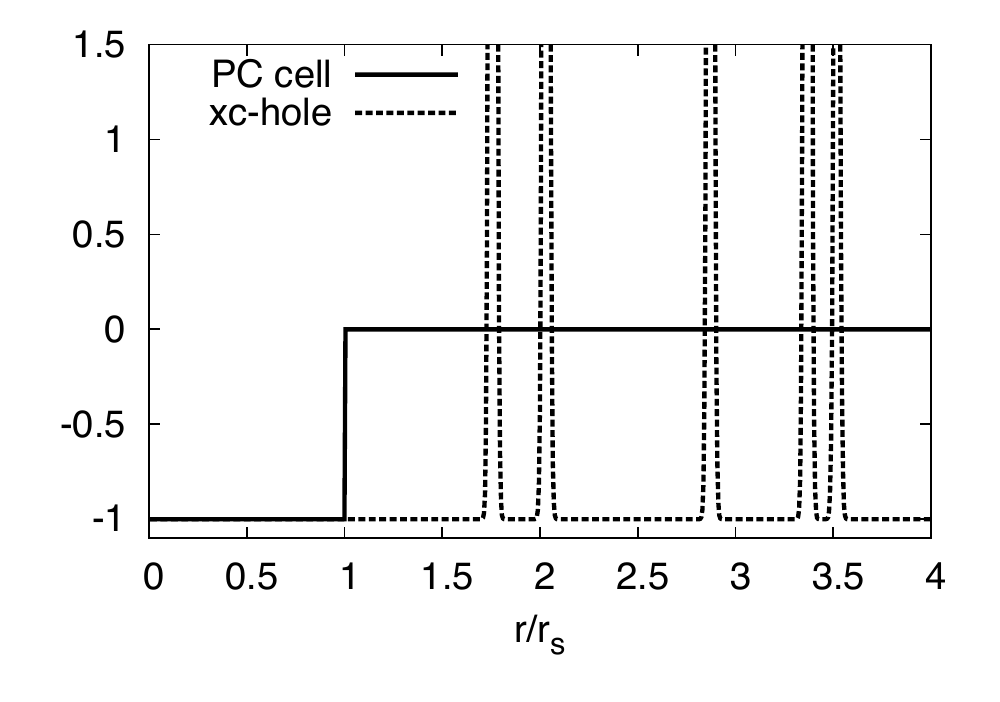}
\caption{PC cell divided by the density $\rho$ (i.e., $c(r)=-\theta(r_s-r)$), and the pair-correlation function $g(r)-1$, corresponding to the exchange-correlation hole divided by the density $\rho$ for the extreme low-density electron gas.}
\label{fig_PCvsxchole}
\end{figure}

In the  case of the uniform electron gas, the  SCE co-motion functions are simply the positions of the bcc lattice points with origin fixed at the reference electron. Notice that the constraint that the density is uniform, Eq.~\eqref{eq_VeeSCEconstr}, forces us to consider a ``floating'' Wigner crystal, which corresponds to the linear superposition of all the possible origins and orientations of the crystal, thus restoring the translational symmetry. The exchange-correlation hole $\rho(g(r)-1)$, with $g(r)$ the pair-distribution function, can then be simply constructed by considering that the expected numbers of electrons in a spherical shell of radius $r$ around the reference electron at the origin is given by
\beq
dN(r|0)=\rho\, g(r)\,4\pi r^2.
\eeq 
We can then place very narrow normalized gaussians (almost delta functions) at the bcc sites around the reference electron and take the spherical average. This way, we obtain the extreme low-density limit of $g(r)$. In Fig.~\ref{fig_PCvsxchole} we compare this low-density (or SCE) $g(r)-1$ with the PC cell $c(r)$ in the same units, $c(r)=-\theta(r_s-r)$, with $\theta(x)$ the Heaviside step function. We see that the two are very different, except for $r/r_s\le 1$. The exchange-correlation hole has positive peaks (indicating the positions of the other electrons) that extend to $r\to\infty$ (perfect long-range order). Notice that the exchange-correlation hole for the broken symmetry solution (without translational invariance) would be, instead, much less structured, but here we are interested in the solution constrained to the uniform density. The way the electrostatic energy is calculated from the PC cell and the exchange-correlation hole is also different:\cite{SeiPerKur-PRA-00}
\begin{eqnarray}
w	& = & \rho \int \frac{g(r)-1}{r}d \rv  \label{eq_engofr} \\
w	& = & -\rho\int \frac{c(r)}{r}d \rv+\frac{\rho^2}{2}\int d\rv\int d\rv'\frac{c(r)c(r')}{|\rv-\rv'|}.
\end{eqnarray}
When we use the exchange-correlation hole to evaluate the energy, Eq.~\eqref{eq_engofr}, we need to evaluate an infinite sum (all the peaks in Fig.~\ref{fig_PCvsxchole}) which converges very badly (the Madelung sum), and that can be dealt with, for example, the Ewald method. When we use the PC cell, instead, we face two very simple, short-ranged, integrals.\cite{Ons-JPC-39} The two results differ only by 0.45\%, as it was already noted in Ref.~\onlinecite{LieNar-JSP-75}, where it was also proven that the PC value is a rigorous lower bound for the energy of the uniform electron gas.

Notice that if, instead, we consider the PC cell as a model for the exchange correlation hole and we use $c(r)$ in Eq.~\eqref{eq_engofr} instead of $g(r)-1$, we get a very poor result\cite{SeiPerKur-PRA-00} with an error of $\sim 17\%$. The PC model does approximate the electrostatic potential of the exchange-correlation hole by constructing it in a different way.

\end{document}